\documentclass[11pt,a4paper]{article}
\pdfoutput=1
\usepackage[english]{babel}
\usepackage[T1]{fontenc}
\usepackage[latin1]{inputenc}
\usepackage{amsfonts,amsbsy,bm,euscript,mathrsfs}
\usepackage{amssymb,stmaryrd,faktor,slashed,arydshln}
\usepackage[x11names]{xcolor}
\usepackage[tbtags]{amsmath}
\usepackage[linktocpage,bookmarks=true,colorlinks=true,linkcolor=black,citecolor=black,urlcolor=black,bookmarksnumbered]{hyperref}

\usepackage[nosort]{cite}

\numberwithin{equation}{section}

\makeatletter
\renewcommand\section{\@startsection {section}{1}{\z@}
{-3.5ex \@plus -1ex \@minus -.2ex}
{2.3ex \@plus.2ex}
{\normalfont\Large\bfseries}}
\renewcommand\subsection{\@startsection{subsection}{2}{\z@}
{-3.25ex\@plus -1ex \@minus -.2ex}
{1.5ex \@plus.2ex}
{\normalfont\large\bfseries}}
\def\maketag@@@#1{\hbox{\m@th\normalfont\normalsize#1}}
\makeatother

\newcommand \arxivlink [1]{\href{http://arxiv.org/abs/#1}{\tt arXiv:#1}}
\newcommand \foot [1] {\footnote{#1\vspace{3pt}}}
\newcommand \rf [1] {(\ref{#1})}

\def \texpdf {\texorpdfstring}
\def \be {\begin{eqnarray}}
\def \ee {\end{eqnarray}}
\def \p {{\partial}}

\def \Tr {{\rm Tr}}

\def \tr {{\rm tr}}

\def \Tr {{\rm Tr}}
\def \ha {{1 \over 2}}
\def \td {\tilde}
\def \ci{\cite}

\def \const {{\rm const}}

\def \z {\zeta}

\def \a {\alpha}
\def \b {\beta}
\def \e {\varepsilon}
\def \p {\phi}

\def \del {\partial}
\def \a {\alpha}

\def \g {\gamma}
\def \s {\sigma}
\def \z {\zeta}

\def \ov {\over}

\def \b {\beta}

\def \det {\hbox{det}}
\def \ci {\cite}

\def \Tr {{\rm Tr}}

\def \const {{\rm const}}

\def \td {\tilde}

\def \m {\mu}

\def \e {\epsilon}

\def \la {\label}

\def \adss {$AdS_5 \times S^5~$}
\def \ov {\over}

\def \ha {{1\ov 2}}

\def \no {\nonumber}

\def \del {\partial}

\def \la {\label}

\def \adss {$AdS_5 \times S^5$\ }

\def \p {\phi}

\def \ov {\over}

\def \varpi {{\rm w}}

\def \ep {\epsilon}

\def \Tr {{\rm Tr}}

\def \s {\sigma}

\def \ha {{{\textstyle{1 \ov2}}}}

\def \eqref {\rf}

\def \adss {$AdS_5\times S^5$\ }

\def \iffa {\iffalse}

\def \diag {{\rm diag}}

\def \Ad {{\rm Ad}}

\def \ed {\end{document}}
\def \tr {{\rm Tr\, }}
\def \sm {$\sigma$-model\ }
\def \sms {$\sigma$-models\ }

\def \A {{\rm A}}
\def \be {\begin{equation}}
\def \ee {\end{equation}}
\def \l {\lambda}

\begin{document}
%%%%%%%%%%%%%%%%%%%%%%%%%%%%%%%%%%%%%%

\iffalse
%%%%%%%%%%%%%%%%%%%%%%%%%%%%%%%%%%%%%%
Title:
Homogeneous Yang-Baxter deformations as non-abelian duals of the AdS_5 sigma-model
Authors:
B. Hoare, A.A. Tseytlin
Abstract:
We propose that the Yang-Baxter deformation of the symmetric space sigma-model
parameterized by an r-matrix solving the homogeneous (classical) Yang-Baxter
equation is equivalent to the non-abelian dual of the undeformed model with
respect to a subgroup determined by the structure of the r-matrix. We
explicitly demonstrate this on numerous examples in the case of the AdS_5
sigma-model. The same should also be true for the full AdS_5 x S^5 supercoset
model, providing an explanation for and generalizing several recent
observations relating homogeneous Yang-Baxter deformations based on non-abelian
r-matrices to the undeformed AdS_5 x S^5 model by a combination of T-dualities
and non-linear coordinate redefinitions. This also includes the special case
of deformations based on abelian r-matrices, which correspond to TsT
transformations: they are equivalent to non-abelian duals of the original model
with respect to a central extension of abelian subalgebras.
Comments:
28 pages; v2: comments and references added
Report Number:
Imperial-TP-AT-2016-03
%%%%%%%%%%%%%%%%%%%%%%%%%%%%%%%%%%%%%%
\fi

\setcounter{equation}{0}
\setcounter{footnote}{0}
\setcounter{section}{0}

\thispagestyle{empty}

{\small \hfill Imperial-TP-AT-2016-03}

\begin{center}
\vspace{1.5truecm}

{\Large \bf Homogeneous Yang-Baxter deformations \\ \vspace{0.2cm} as non-abelian duals of the $\mathbf{AdS_5}$ $\mathbf{\sigma}$-model}

\vspace{1.5truecm}

{B. Hoare$^{a,}$\footnote{bhoare@ethz.ch} and A.A. Tseytlin$^{b,}$\footnote{Also at Lebedev Institute, Moscow. tseytlin@imperial.ac.uk }}\\

\vspace{0.5truecm}

{\em $^{a}$ Institut f\"ur Theoretische Physik, ETH Z\"urich,\\ Wolfgang-Pauli-Strasse 27, 8093 Z\"urich, Switzerland.}

\vspace{0.1truecm}

{\em $^{b}$ The Blackett Laboratory, Imperial College, London SW7 2AZ, U.K.}

\vspace{1.0truecm}
\end{center}

\begin{abstract}
We propose that the Yang-Baxter deformation of the symmetric space
$\sigma$-model parameterized by an $r$-matrix solving the homogeneous
(classical) Yang-Baxter equation is equivalent to the non-abelian dual of the
undeformed model with respect to a subgroup determined by the structure of the
$r$-matrix. We explicitly demonstrate this on numerous examples in the case of
the $AdS_5$ $\sigma$-model. The same should also be true for the full $AdS_5
\times S^5$ supercoset model, providing an explanation for and generalizing
several recent observations relating homogeneous Yang-Baxter deformations based
on non-abelian $r$-matrices to the undeformed $AdS_5 \times S^5$ model by a
combination of T-dualities and non-linear coordinate redefinitions. This also
includes the special case of deformations based on abelian $r$-matrices, which
correspond to TsT transformations: they are equivalent to non-abelian duals of
the original model with respect to a central extension of abelian subalgebras.
\end{abstract}

\newpage
\setcounter{equation}{0}
\setcounter{footnote}{0}
\setcounter{section}{0}

%%%%%%%%%%%%%%%%%%%%%%%%%%%%%%%%%%%%%%
\tableofcontents
%%%%%%%%%%%%%%%%%%%%%%%%%%%%%%%%%%%%%%

%%%%%%%%%%%%%%%%%%%%%%%%%%%%%%%%%%%%%%
\section{Introduction}\label{secint}
%%%%%%%%%%%%%%%%%%%%%%%%%%%%%%%%%%%%%%

The class of integrable deformations of the \adss superstring model based on
the ``Yang-Baxter'' \sm \cite{Klimcik:2002zj,Klimcik:2008eq,dmv} has recently
been under active investigation. It generalizes the usual group-space or
coset-space \sm as $\mathcal{L}=\Tr ( J J) \to \Tr ( J \mathcal{O} J)$ where
$J= g^{-1} d g$ and $\mathcal{O}$ depends on $g$ and a constant antisymmetric
operator $R$ acting on the Lie algebra (referred to as the ``$r$-matrix'')
satisfying a (modified) classical Yang-Baxter equation (cYBE), i.e. $[RX,RY] -
R ( [RX,Y] + [X, RY]) = c [X,Y] $.

The two inequivalent cases are $c=1$ and $c=0$.\foot{One may also consider
$c=-1$. However, no solutions exist for the real form of the superalgebra
$\mathfrak{psu}(2,2|4)$, the symmetry algebra of the $AdS_5 \times S^5$
superstring.} The first case represents a non-trivial $q$-deformation of the
symmetry algebra of the original symmetric space \sm \cite{Delduc:2013fga,dmv}.
The second case (based on the homogeneous cYBE) studied in
\cite{yabe1,yabe2,yjor,ysum,vtsum,ysugra1,Hoare:2016hwh,ysugra2,Borsato:2016ose,Osten:2016dvf}
appears to be simpler and more closely related to the original coset
$\s$-model. Indeed, it was observed on particular examples
\cite{yabe1,yabe2,yjor,vtsum} and proved in general \cite{Osten:2016dvf} that
for abelian $r$-matrices the resulting deformed model can be obtained from the
original undeformed one by abelian T-dualities (more precisely, by a TsT
transformation combining T-duality with a linear coordinate shift which is a
special case of the $O(d,d)$ T-duality transformation).

Furthermore, for several examples defined by non-abelian $r$-matrices solving
the homogeneous cYBE it was recently observed \cite{ysugra2,Borsato:2016ose}
that the resulting deformed \sm can be related to the original \adss supercoset
model by a combination of T-dualities along {\it non}-commuting directions and
{\it non}-linear coordinate transformations (required to be able to perform the
T-dualities).

Here we will generalize (and provide an explanation for) these observations by
demonstrating that the homogeneous YB deformations of a symmetric space \sm are
equivalent to non-abelian duality (NAD) transformations
\cite{nadold,os,vene,gr,alv,elit} of the original model with respect to various
(in general, non-semi-simple) subgroups of the global symmetry group.
We will focus on the bosonic $AdS_5$ \sm but the same
should be true in general and also for the supercoset \adss model.

The subgroup in which we will dualize is determined by the structure of the
$r$-matrix. Not all subgroups have a corresponding classical $r$-matrix and
hence not all NAD transformations correspond to a homogeneous YB deformation.
Indeed the corresponding subalgebra should be quasi-Frobenius (or a central
extension thereof), which, in particular, implies it is solvable. Therefore, the
correspondence is absent, for example, if the subgroup is non-abelian and
simple (such as $SO(1,2) \subset SO(2,4)$ or $SO(3) \subset SO(6)$). In these
cases one would not expect the NAD to be a deformation of the original
$\sigma$-model. On the other hand, the case of abelian $r$-matrices is
naturally included: as we shall explain below, the abelian TsT transformation
may be viewed as a special case of non-abelian duality with respect to a
central extension of an abelian symmetry group. It is an interesting open
question what the criteria is for a subgroup to give NAD model that can be
understood as a deformation of the original $\sigma$-model, when one needs
to centrally extend the group, and if these cases are in correspondence with
the homogeneous YB deformations.

%v2
As the non-abelian duality should preserve the classical integrability of a
symmetric \sm \cite{sfet}, one may then, instead of studying homogeneous YB
deformations, directly consider all possible NAD transforms of the \adss \sm
with respect to all possible (centrally-extended) subgroups of $PSU(2,2|4)$.
As mentioned above, there will
%v3
be  special NAD models that have a
``deformation'' interpretation (i.e.   depend  on  free parameters that  when
taken to zero give back the original model) and other NAD models
(corresponding to simple subgroups) that are close cousins of the original
model (e.g.  sharing the same first-order structure) but not reducing to it in
a limit and  not allowing one to reverse the NAD transform.\foot{A  potentially
important feature of the NAD  transform  compared to the YB   deformation is that
it may   be applied also to the  less symmetric (and non-integrable) D3-brane
background away from the near-horizon limit. As in the case of the TsT duality
example in \ci{Maldacena:1999mh} this may help to understand the dual gauge theory
interpretation  of the resulting model.} 

An  advantage of NAD over the YB deformation is that it can be performed
systematically as a path integral transformation (determining also the 
%v3
dilaton).
This allows us, in particular, to answer the question of which NAD
transforms of the \adss \sm will still be Weyl-invariant so that the
corresponding backgrounds will be solutions of the standard type II
supergravity equations (and will thus define critical string models).

In general, the NAD transform of a Weyl-invariant \sm is not Weyl but only
scale-invariant \cite{elit} (in the context of GS superstring \sm this means
that the corresponding dual background solves only the generalized supergravity
equations of \cite{Arutyunov:2015mqj,Wulff:2016tju}). The condition for NAD to
preserve Weyl invariance is that the
structure constants of the group which is dualized should
be traceless, i.e.
$n_a \equiv f^{c}_{\ c a} =0$ \cite{alv,elit}.
An equivalent ``unimodularity'' condition was found also for
the homogeneous YB deformations of the \adss \sm as a condition on the
corresponding non-abelian $r$-matrices \cite{Borsato:2016ose}. If $n_a\not=0$
the NAD-transformed background does not solve the Weyl-invariance or
supergravity equations but it can still be mapped by a formal T-duality (along
the $n_a$ direction, as discussed in section \ref{secnad}) to a proper
supergravity solution, in agreement with the general discussion in
\cite{Arutyunov:2015mqj}.

While abelian TsT transformations (and thus equivalent abelian YB deformations)
of \adss are related, via AdS/CFT, to non-commutative gauge theories
\cite{Hashimoto:1999ut,Maldacena:1999mh,Lunin:2005jy,Dhokarh:2008ki,vtsum} the
role of the non-abelian duals of \adss (beyond a supergravity solution
generating technique) is presently unclear. In the TsT case one is guided by
the action of T-duality on open strings or D-branes but similar intuition is
absent in the NAD case (cf. though \cite{Forste:1996hy}). Still, the quantum
Weyl-invariant and integrable \sms obtained by NAD from the \adss model may
provide new non-trivial examples of solvable string models.

Let us briefly comment on the case of the {\it in}homogeneous YB deformation of
the \adss \sm \cite{dmv} that non-trivially depends on one ``quantum''
deformation parameter. The corresponding 2d theory is scale but not Weyl
invariant, i.e. the associated background \cite{ABF,HT2} only solves the
generalized equations of \cite{Arutyunov:2015mqj,Wulff:2016tju}. At the same
time, it is classically related \cite{Vic,HT1}, by Poisson-Lie (PL) duality
\cite{KS}, to the ``$\l$-model'' of \cite{hms2,hms1} (generalizing the bosonic
model of \cite{sfet}) which is Weyl-invariant at the quantum level
\cite{apphol,BTW,Lunin,Borsato:2016ose}. This suggests that as for the
non-abelian duality in the case of $n_a \not=0$ the PL duality (which should
preserve quantum equivalence on a flat 2d background and thus scale invariance
\cite{Alekseev:1995ym}) here should be ``Weyl-anomalous'' at the quantum level.

A special ``undeformed'' (level $k\to \infty$ or $q=e^{i\pi \ov k} \to 1$)
limit of the $\l$-model is just the NAD of the \adss supercoset model with
respect to the full $PSU(2,2|4)$ which is still Weyl invariant. The $\l$-model
can thus be interpreted as a Weyl-invariance preserving $q$-deformation of the
full NAD of the \adss superstring model. It would be of interest to see if one
can construct similar non-trivial deformations of NADs of \adss with respect to
some subgroups of $PSU(2,2|4)$ (cf. \cite{sfth}).

Among other open questions, it would be useful to give a general proof of the
equivalence between the homogeneous YB deformations of the \adss supercoset
model and its non-abelian duals with respect to the corresponding subalgebras.
One may be able to establish this relation by identifying the underlying
first-order actions which include auxiliary gauge fields.\foot{For the bosonic
YB deformed model such a first-order action was given in \cite{HT1}.}

The structure of this paper is as follows. We shall start in section 2 with a
review of the non-abelian duality transformation of a bosonic $\s$-model,
explaining also how one can interpret the abelian TsT transformation as a
special case. In section 3, after reviewing the Yang-Baxter deformation and
the NAD of the symmetric space \sm we shall turn our attention to the $AdS_5$
$\sigma$-model. For a range of cases of different types of $r$-matrices
(abelian and non-abelian, both unimodular and jordanian) we will explicitly
demonstrate that the Yang-Baxter deformation is equivalent to the non-abelian
dual of the $AdS_5$ \sm with respect to a specific (centrally-extended)
subalgebra that can be identified from the $r$-matrix. In the Appendix we will
present a large number of additional jordanian examples.

%%%%%%%%%%%%%%%%%%%%%%%%%%%%%%%%%%%%%%
\section{Non-abelian duality}\label{secnad}
%%%%%%%%%%%%%%%%%%%%%%%%%%%%%%%%%%%%%%

In this section we shall discuss non-abelian duality (NAD) in bosonic string
\sms \cite{nadold,os,vene,gr,alv,elit,SF}. NAD relates one bosonic \sm with a
non-abelian global symmetry group $H$ (i.e. with $H$-invariant metric,
$B$-field and dilaton $\p$) to another one that generically has a smaller
symmetry group (and no symmetry in the case when $H$ is non-abelian and
simple). The standard abelian (or ``torus'') T-duality \cite{bu} is a
special case when $H$ is abelian, i.e. $\mathbb{R}^d$ or $U(1)^d$. The
$O(d,d)$ generalization of the T-duality (and, in particular, the special TsT
case) can also be viewed, as we shall explain below, as a particular case of
NAD when one considers a central extension of the abelian group.

If the original \sm is classically integrable then NAD maps it, as for the
standard T-duality case 
%v3
(see, e.g., \cite{Frolov:2005ty,Hatsuda:2006ts,Ricci:2007eq}),
 to a classically integrable model
\cite{sfet}. If the original \sm is quantum Weyl-invariant, its NAD counterpart
is also Weyl-invariant unless $H$ is non-semisimple with the generators in the
adjoint representation satisfying $\tr T_a\not=0$, i.e. $n_a\equiv f^c_{\
ca}\not=0$ \cite{alv,elit}. In the latter case the NAD transformation preserves
only the scale invariance of the $\s$-model.

It should be possible to generalize NAD to the case of the GS superstring \sm
and then for $n_a =0$ NAD should again preserve Weyl invariance, i.e. (viewed
as a transformation on the target space couplings of the $\sigma$-model) will
map a supergravity solution to a supergravity solution.\foot{A prescription for
the NAD transformation of the RR fields has already been proposed (by analogy
with T-duality) in \cite{sfth2}. It has also been checked on examples that for
simple groups $H$, like $SU(2)$, NAD indeed maps supergravity solutions to
supergravity solutions and that the transformation is a symmetry of the %v3
supersymmetry variations of supergravity \cite{Lozano:2011kb}.} For $n_a \not=0$
NAD may not preserve Weyl invariance,\foot{Here we are not considering the
possibility of dualizing in fermionic directions.} i.e. may transform a
Weyl-invariant \sm into a scale-invariant one. Then the original target space
background that was a supergravity solution will be mapped by NAD to a solution
of the generalized supergravity equations or superstring \sm scale invariance
conditions \cite{Arutyunov:2015mqj,Wulff:2016tju}.

As we shall see below, in agreement with the general expectation
\cite{Arutyunov:2015mqj}, in that case one can still apply a formal (classical,
i.e. ignoring the dilaton transformation) T-duality to associate to the
resulting generalized background a solution of the standard Weyl-invariance or
supergravity equations that has a linear non-isometric term in the dilaton
obstructing the reverse T-duality at  the  quantum level \cite{HT1,HT2}.

%%%%%%%%%%%%%%%%%%%%%%%%%%%%%%%%%%
\subsection{Non-abelian duality of bosonic \texpdf{$\sigma$-model}{sigma-model} in curved 2d background}\label{ssecnad}
%%%%%%%%%%%%%%%%%%%%%%%%%%%%%%%%%%

Let us start with a bosonic string \sm depending on group $H$ coordinates $h$
and ``spectator'' coordinates $x^r$ that is invariant under a global
$H$-symmetry, $h \to h_0 h, \ h_0\in H$. Its action can be written in conformal
gauge $g_{ij}= e^{2 \s} \eta_{ij}$, with $\eta = \diag(-1,1)$, as
\begin{align}
I[h,x] & = {1\ov 4 \pi \a'} \int d^2 z \big[ E_{ab}(x)\, J_+^a J^b_- + L_{ar}(x)\, J_+^a j^r_ - + M_{sb}(x)\, j_+^s J^b_ - + K_{rs}(x)\, j_+^r j^s_ -
\no \\
& \hspace{75pt} - 2 \a' \p(x)\, \del_+ \del_- \s
\big] \ , \la{1} \\
J^a_i & = \tr (h^{-1} \del_i h \, \bar T^a ) \ , \qquad j^r_i = \del_i x^r \ , \qquad [T_a,T_b]=f^c_{\ ab}T_c \ , \qquad
\tr(T_a \bar{T}^b)= \delta_{a}^b \la{2} \ .
\end{align}
Here $i=0,1$, $\del_\pm = \del_0 \pm \del_1$, \ $a,b= 1, 2, \ldots, d=\dim H$,
\ $r,s= 1,\ldots, n$.
$T_a$ is a basis for the algebra $\mathfrak{h}$ of $H$
(taken below to be given by matrices in the adjoint representation, $T_a = - f^c_{\ ab}$)
and $\bar T^a$ is a dual algebra basis (assuming $H$ is embedded into a
simple group).\foot{For example, the case of a coset space may be included by
taking $E_{ab}$ to be degenerate.}
We have also included a possible $H$-invariant dilaton term ${1\ov 4 \pi} \int d^2 z
\sqrt g R^{(2)} \phi(x) = -{1\ov 2 \pi} \int d^2 z\, \p(x)\, \del^2 \s$
where $\s$ is the conformal factor of the 2d metric. The path integral over
$h$ with the action \rf{1} can be obtained from the path integral over the 2d
gauge field $A_\pm^a$ and ``Lagrange-multiplier'' field $v_a$ with the
following first-order or ``interpolating'' action
\begin{align}
\hat I[A,v,x] & = {1\ov 4 \pi \a'} \int d^2 z \big[ E_{ab}(x)\, A_+^a A^b_- + L_{ar}(x)\, A_+^a j^r_ - + M_{sb}(x)\, j_+^s A^b_ - + K_{rs}(x)\, j_+^r j^s_ - \no \\
& \hspace{75pt}
+\ v_a \ep^{ij} F^a_{ ij} + 2 \a' \s\, n_a \del^i A^a_i - 2 \a' \p(x)\, \del_+ \del_- \s \big] \la{3} \ , \\
& \la{4} F^a_{ij} \equiv \del_i A^a_j - \del_j A^a_i + f^a_{\ bc} A^b_i A^c_j \ ,\qquad \qquad n_a = \tr T_a= f^c_{\ c a} \ .
\end{align}
Integrating over $v_a$ gives $F_{ij} =0$ or $A_i= A^a_i T_a = h^{-1} \del_i h $
and the first line in \rf{3} becomes equivalent to the first line of \rf{1} on
a flat 2d background.

The $n_a$-dependent term \cite{elit} in the second line of \rf{3} \cite{elit},
which is non-local in the 2d metric ($\sigma = - \ha \del^{-2} \sqrt g R^{(2)}
$), is important for the quantum equivalence of \rf{3} and \rf{1} on a curved
2d background. It comes from the Jacobian of the transformation from $A_+ =
h^{-1} \del_+ h, \ A_-=h'^{-1} \del_- h'$ to $h, h'$ on a curved 2d background
and is related to the ``mixed'' anomaly \cite{alv} ($\del_i {\delta
\Gamma \ov \delta A_i} \sim R^{(2)}$) which is present only if $H$ is such that
$n_a \not=0$.

The dual of the model \rf{1} is then found by integrating out the gauge field $A^a_i$ in \rf{3}
\begin{align}
\td I[v,x] & = \int d^2 z \Big[\big(\del_+ v_a + M_{sa}(x)\, \del_+ x^s + \a' n_a \del_+ \s\big) N^{ab} \big(\del_- v_b - L_{br}(x)\, \del_- x^r - \a' n_b \del_- \s\big) \no \\
& \hspace{75pt} \la{5}
+ K_{rs}(x)\, \del_+x^r \del_- x^s - 2 \a' \big( \p(x) + \ha \ln \det N \big) \, \del_+ \del_- \s \Big] \ ,
\\
N^{ab} & = \big[ E_{ab}(x) - v_c f^c_{\ ab} \big]^{-1} \la{6} \ .
\end{align}
If $n_a \not=0$ \rf{5} does not have the interpretation of a local \sm action
on a curved 2d background. If we formally ignore the $n_a$-dependent terms in
\rf{5} the corresponding dual target space background (metric $\td G$,
2-form field $\td B$ and dilaton $\td \p$) will no longer represent a solution of the
critical string Weyl-invariance conditions \cite{elit}, explaining what was
observed in \cite{vene} for a specific example. Considered on a flat 2d
background the dual \sm \rf{5} will still be scale invariant so that the dual
metric and $B$-field will solve the scale-invariance conditions (i.e., in the
superstring context, the generalized supergravity equations
\cite{Arutyunov:2015mqj,Wulff:2016tju}).

Let us note that the scale-invariant dual background $(\td G, \td B)$ has a remarkable property that it can still
be naturally associated to a proper
solution of the Weyl-invariance
conditions by applying a formal (2d flat space) duality transformation in the direction of the
isometric part of $v_a$ parallel to $n_a$.
The key observation is that while for $n_a\not=0$ the model \rf{5} can not be interpreted as
a local \sm on a curved 2d background, applying a 2d duality transformation
to one scalar field in \rf{5} restores such an
interpretation.\foot{Equivalently, one may say that in order to preserve the
local \sm interpretation of the dual model and thus its Weyl invariance
one should not dualize in the $n_a$-direction of the algebra in the first place.}
Let us split the Lagrange multiplier field $v_a$ in \rf{3} as
$ v_a = u_a + n_a y $. Then $y$ will appear in \rf{3} and thus also in \rf{5} only through its derivatives, i.e.
shifts of $y$ will be an abelian isometry of the
dual background.\foot{If
$H$ is non-abelian and simple (such that $n_a=0$) then the dual background has no isometries.
If $n_a \neq 0$ then the dual background will have at least one abelian isometry.}
Indeed, we will have
$v_a F^a_{ ij} = u_a F^a_{ ij} + y (\del_i \A_j - \del_j \A_i)$ where $\A_i = n_a A^a_i = \tr A_i $
and we have used that $n_a f^a_{\ bc} = f^d_{\ da} f^a_{\ bc} =0$ as follows from the Jacobi identity.
Then the $y$-dependent and related terms in \rf{3} will be (up to integration by parts)
\be \la{7} v_a \ep^{ij} F^a_{ ij} +2 \a' \s\, n_a \del^i A^a_i
\to -2 \ep^{ij} \del_i y \A_j
+2 \a' \s\, \del^i \A_i \ . \ee
Applying T-duality in the $y$ direction amounts to replacing $\del_i y $ by $B_i$ adding at the same time
the term
$-2 \ep^{ij} \del_i \td y\, B_j$. Then integrating over $B_i$ gives
$\A_i= \del_i \td y$
and thus the last term in \rf{7} becomes (after integrating by parts)
$2 \a' \s\, \del^i \A_i
\to -2 \a' \td y\ \del_+ \del_- \s\ $. The latter
is the standard $R^{(2)} \p$ dilaton term with $\p$ linear in the dual coordinate,
$\p\sim \td y$.
Integrating the rest of $A_i^a$ out will then give a local Weyl-invariant \sm with
a linear non-isometric dilaton term.
The above steps can be carried out explicitly, e.g., on the examples of the models discussed in \cite{elit}.

We conclude that the scale-invariant \sm background found by applying the NAD
transformation to the action \rf{1}, which, taking $n_a \neq 0$, does not solve
the Weyl-invariance equations, is still related by formal T-duality to a
Weyl-invariant background with a linear non-isometric dilaton. This background cannot then be
T-dualized back to give a standard Weyl-invariant \sm with local couplings
to the worldsheet metric. Thus NAD provides a particular example of the
general case discussed in \cite{Arutyunov:2015mqj}.

%%%%%%%%%%%%%%%%%%%%%%%%%%%%%%%%%%%%%%%
\subsection{TsT duality as special case of non-abelian duality}\label{sectst}
%%%%%%%%%%%%%%%%%%%%%%%%%%%%%%%%%%%%%%%

Let us now consider the particular case of $O(2,2)$ T-duality where one starts with
a metric having two abelian isometries, $ds^2 = dx^2 + f_1(x) dy_1^2 + f_2 (x) dy_2^2$,
applies T-duality $y_1 \to \td y_1$, then shifts $y_2 \to y_2 + \g \td y_1$ introducing the parameter $\g$, and finally T-dualizes
back $\td y_1 \to {\td {\td y}}_1 \equiv y_1$. This generates the following non-trivial background depending on $\g$
\begin{equation}\begin{split} \la{8}
ds^2 & = dx^2 + U(x) \big[ f_1(x) dy_1^2 + f_2 (x) dy_2^2\big] \ ,
\\ B_{y_1y_2} & = \g f_1 f_2 U \ , \qquad e^{2\p} = U(x) \equiv {1 \ov 1 + \g^2 f_1(x) f_2(x) } \ . \end{split}\end{equation}
Applications of such TsT  transformations were discussed, e.g., in \cite{tst,Hashimoto:1999ut,Maldacena:1999mh,Lunin:2005jy,Frolov:2005ty}.

Below we shall show that this transformation of the corresponding 2d \sm may be viewed as a special case of NAD
with the algebra of $H$ being the centrally-extended 2d translation algebra (or Heisenberg algebra)
\be \la{9}
[P_r, P_s]= \ep_{rs} Z\ , \qquad [P_r,Z]=0\ , \qquad r,s=1,2 \ . \ee
Let us start with $\mathcal{L}= f_1 ( \del_i y_1)^2 + f_2 (\del_i y_2)^2$.
The ``interpolating'' action that corresponds to the first two steps of the above TsT transformation may be written as
\be \la{10}
\td{\mathcal{L}} = f_1 (A_i)^2 + f_2 (\del_i y_2 + \g \del_i \td y_1)^2 + 2 \ep^{ij} \del_i \td y_1 A_j \ .\ee
T-dualizing again $\td y_1\to y_1$ by introducing another abelian gauge field $A'_i$ we find
\be \la{11}
\mathcal{L}'= f_1 (A_i)^2 + f_2 (\del_i y_2 + \g A'_i )^2 + 2 \ep^{ij} A'_i A_j + 2 \ep^{ij} \del_i y_1 A'_j \ .\ee
Redefining $A_i \equiv A^{1}_i$,  \ $\del_i y_2 + \g A'_i \equiv A^{2}_i$, %v2
and then sending $y_1 \to -\g v_2$, \ $y_2 \to \gamma v_1$ we arrive at
\be \la{12}
\mathcal{L}'= f_1 (A^{1}_i)^2 + f_2 (A^{2}_i)^2
+2 \ep^{ij} \big( v_1 \del_i A^{1}_j + v_2 \del_i A^{2}_j \big)
- 2\g^{-1} \ep^{ij} A^{1}_i A^{2}_j \ .\ee
If we first integrate over $v_r$ we have $A^{r}_i=\del_i y^r_i$ and thus go
back to the original model (the last term in \rf{12} is then total derivative).
Integrating instead over the two gauge fields in \rf{12} one ends up with the
\sm corresponding to the TsT background \rf{8} (with $v_r \to y_r$). Note that without the last
$-2\g^{-1} \ep^{ij} A^{1}_i A^{2}_j $ term \rf{12} is the first-order action for
T-dualizing on both $y_1$ and $y_2$.\foot{Indeed, the $\g=\infty$ limit of
\rf{8} is the double T-dual background. This last term in \rf{10} which is
absent in this limit, is analogous to ``current-current'' deformation related
to $O(2,2)$ duality (cf. \cite{Kiritsis:1993ju}).}

Let us now derive the same Lagrangian \rf{12} starting instead with the NAD
model \rf{3} with
$\mathfrak{h}$ taken to be the centrally-extended algebra
\rf{9} with $T_a=(P_1,P_2, Z)$ (for which only $f^3_{\ 12}$ is non-zero and
thus $n_a=0$). Here
\be \la{13}
v_a \e^{ij} F_{ij}^a = 2 \e^{ij} \big[ v_1 \del_i A^1_j + v_2 \del_i A^2_j + w ( \del_i C_j + A^1_i A^2_j) \big]
\ , \ee
where $w\equiv v^3$ and $C_i\equiv A^3_i$ corresponds to the central generator
$Z$ in \rf{9}. Note that as the original \sm is assumed to be invariant under the
abelian translations, the central generator $Z$ should act trivially on the
coordinates. Gauging it is still possible as integrating out $v_r$ and $w$
brings us back to the original model: we get $A^r_i = \del_i y^r $ (and $dC $
expressed in terms of $y^r$), so that $ f_r (A^{r}_i)^2 \to f_r (\del_i
y^r)^2$.

Since the gauge field $C_i$ does not enter the rest of the \sm action, we can
readily integrate it out getting the condition $w=w^{(0)}=\const$. Then \rf{13} becomes
the same as the last three terms in \rf{12} with $w^{(0)}\sim -\g^{-1}$. %v2

Thus considering the central extension of the abelian translation group allows
one to introduce an extra free parameter $\g$ (absent in the standard first-order abelian
T-duality action): the TsT parameter $\g$ acquires the
interpretation of the background value of the dual coordinate corresponding to
the central generator $Z$.
Let us note also that in the example we considered above
the origin of the $B$-field of
the resulting background can be traced to the non-abelian nature of the
Heisenberg algebra \rf{9} (which is also related to the non-commutativity of the
dual gauge theory \cite{Hashimoto:1999ut,Maldacena:1999mh,Lunin:2005jy}).

%v2
One can obviously consider various generalizations. 
First, one can readily   repeat the above discussion for generic model with two abelian isometries
$\mathcal{L}  =  f_1 ( \del_i y_1)^2 + f_2 (\del_i y_2)^2
+ g_1 \partial^i y_1 \partial_i y_2 + g_2 \ep^{ij} \partial_i y_1 \partial_j y_2 + h^{i}_1 \partial_i y_1 +h^{i}_2 \partial_i y_2 $
where $f_{r}$, $g_{r}$ and  $h^{i}_r$  are functions of the remaining fields, including fermionic degrees of freedom.
If the rank of the abelian
isometry algebra is greater than 2, constructing its central extensions and
performing NAD will introduce several continuous parameters
$w^{(0)}_m$ as expected in the general $O(d,d)$ T-duality case.
Instead of an abelian isometry group one may also
have a non-abelian one, $H$, that has an abelian (translational) subgroup. While
the direct application of the NAD transformation with respect to
$H$ may give a dual model with no free parameters, starting instead with a
centrally-extended group and then applying NAD will lead to a model
containing several free parameters. In general, the central extension of an
algebra satisfying $f^{c}_{\ ca} = 0$ will also satisfy this property.

For example, one may start with the euclidean $AdS_3$ space $ds^2 =z^{-2} [
dz^2 + dy_1^2 + dy_2^2] $ and perform NAD with respect to the 2d Euclidean
group $ISO(2) \subset SO(1,3)$ with the algebra $[J, P_r] = \ep_{rs} P_s$,
$[P_r,P_s]=0$. Considering its central extension then allows one to introduce
a free parameter. For rank four and higher algebras one may be able to
introduce several parameters. Such examples will be discussed below for the
$AdS_5$ $\sigma$-model.

%%%%%%%%%%%%%%%%%%%%%%%%%%%%%%%%%%%%%%
\section{Homogeneous Yang-Baxter deformations of the \texpdf{$AdS_5$ \sm}{AdS5 sigma-model} and non-abelian duality}\label{secgen}
%%%%%%%%%%%%%%%%%%%%%%%%%%%%%%%%%%%%%%

We now turn to demonstrating the conjectured equivalence
between homogeneous Yang-Baxter (YB)
deformations of a coset \sm and non-abelian duals (NADs) of that same model.
We shall first make some general remarks and
then focus on the bosonic $AdS_5$ model but similar considerations should apply
also to the full \adss supercoset model.

In general, the homogeneous YB deformation of the $F/G$ symmetric space \sm is
based on a solution to the classical Yang-Baxter equation (cYBE) for
$\mathfrak{f} = Lie(F)$. These solutions are in correspondence with the
quasi-Frobenius subalgebras of $\mathfrak{f}$.

The homogeneous YB deformations with
$r$-matrices corresponding to abelian subgroups
are equivalent to TsT transformations of the original coset model \cite{Osten:2016dvf}.
As we have seen in section \ref{sectst}, the TsT transformation can, in fact, be
reformulated as the NAD with respect to a centrally-extended abelian subgroup.

Furthermore, in \cite{Borsato:2016ose} it was shown
that the YB deformation of the \adss supercoset model \cite{yjor,dmv}
corresponds to a supergravity solution (i.e. is one-loop Weyl invariant when supplemented with an appropriate dilaton)
if the $r$-matrix $r = r^{\a\b} e_\a \wedge e_\b$
is unimodular, i.e. satisfies
\begin{equation}\label{uni}
r^{\a\b}[e_\a,e_\b] = 0 \ .
\end{equation}
Here the $e_\a$ generate the quasi-Frobenius subalgebra $\mathfrak{h}$ of $\mathfrak{f}$.\foot{Compared
to section \ref{secnad} we now
use Greek indices $\a,\b,\ldots$ for the subalgebra $\mathfrak{h}$ in which we will dualize
and Latin indices $a,b,\ldots$ for the algebra $\mathfrak{f}$.}
The unimodularity condition \eqref{uni}, combined with the quasi-Frobenius property,
implies that the structure constants on $\mathfrak{h}$ satisfy
\cite{Borsato:2016ose}
\begin{equation}
f^{\gamma}_{\ \gamma\alpha} = 0 \ .
\end{equation}
As discussed in section \ref{secnad}, this is precisely the same as the requirement
on $\mathfrak{h}$ for the NAD with respect to $\mathfrak{h}$ of an $H$-invariant \sm to preserve the one-loop
Weyl invariance \cite{alv,elit}.

Based on these observations we shall make a conjecture that the homogeneous YB
deformation based on a classical $r$-matrix is always equivalent to the NAD
transformation of the original coset model with respect to the corresponding
(centrally-extended) quasi-Frobenius subalgebra. As already mentioned, for
abelian $r$-matrices this follows from the results in section \ref{sectst}
and \cite{Osten:2016dvf}. For non-abelian $r$-matrices we will not prove this
conjecture directly, but will provide a comprehensive range of examples
(including those in \cite{Hoare:2016hwh,ysugra2,Borsato:2016ose}), explicitly demonstrating
its validity.

As our primary focus is on deformations of the $AdS_5 \times S^5$ superstring
and its lower-dimensional counterparts here we will restrict our attention to
the bosonic $AdS_5$ $\s$-model. The isometry algebra $\mathfrak{so}(2,4)$
admits a number of different subgroups corresponding to non-abelian
$r$-matrices of various types, both unimodular and jordanian.\foot{Due to the
compact nature of the isometry algebra $\mathfrak{so}(6)$ of $S^5$, it only
admits abelian solutions of the cYBE.}
While it is unclear if it is a universal rule, we
find that for unimodular $r$-matrices we need to consider the
centrally-extended algebra, while for jordanian $r$-matrices this appears not
to be required.

%%%%%%%%%%%%%%%%%%%%%%%%%%%%%%%%%%%%%%%
\subsection{YB deformation and non-abelian duality for the symmetric space \texpdf{\sm}{sigma-model}}
%%%%%%%%%%%%%%%%%%%%%%%%%%%%%%%%%%%%%%%

Our starting point for both the homogeneous YB deformation and the NAD transformation will be
the symmetric space $\s$-model.
Here we will define them for a generic
symmetric space $F/G$ with both $F$ and $G$ being semi-simple. In the case of $AdS_5$ we have
$F = SO(2,4)$ and $G = SO(1,4)$.

For $\mathfrak{f} = Lie(F)$ and $\mathfrak{g} = Lie(G)$ we use the standard bilinear
form on $\mathfrak{f}$ to define $\mathfrak{p}$ as the orthogonal complement of $\mathfrak{g}$
in $\mathfrak{f}$ so that for a symmetric space
\begin{equation}
\mathfrak{f} = \mathfrak{g} \oplus \mathfrak{p} \ , \qquad \Tr[\mathfrak{g}\mathfrak{p}] = 0 \ , \qquad
[\mathfrak{g}, \mathfrak{g}] \subset \mathfrak{g} \ , \qquad
[\mathfrak{g}, \mathfrak{p}] \subset \mathfrak{p} \ , \qquad
[\mathfrak{p}, \mathfrak{p}] \subset \mathfrak{g} \ . \qquad
\end{equation}
The Lagrangian for the symmetric space \sm is then given by
\begin{equation}\label{ssslag}
\mathcal{L} = \Tr[J_+ P J_-] \ , \qquad \qquad J = f^{-1} df \ , \qquad f \in F \ ,
\end{equation}
where $P$ is the projector onto $\mathfrak{p}$,
$\Tr$ is appropriately normalized and $\pm$ are light-cone coordinates on the worldsheet.
This action has a global $F$ symmetry $f \to f_0 f$ and a local $G$ gauge symmetry $f \to f g$.

The homogeneous YB deformation
\cite{Klimcik:2002zj,Klimcik:2008eq,Delduc:2013fga,yjor} of the
symmetric space \sm is defined as
\begin{equation}\label{yblag}
\mathcal{L} = \Tr[J_+ P \frac{1}{1-R_f P} J_-] \ , \qquad\qquad R_f = \Ad_f^{-1} R \Ad_f \ ,
\end{equation}
where operator $R$ is an antisymmetric solution of the cYBE for the algebra $\mathfrak{f}$
\begin{equation}
[RX,RY] = R([RX,Y] + [X,RY]) \ , \qquad X,Y \in \mathfrak{f} \ .
\end{equation}
As this equation is homogeneous in $R$ we can always multiply any solution by an overall constant $\eta$.
Then in the limit $\eta\to 0$
we recover the Lagrangian \eqref{ssslag} of the symmetric space
$\s$-model. The deformed action \eqref{yblag} preserves the local $G$ gauge
symmetry $f \to fg$; however, the global symmetry is broken to a subgroup 
%v3
of $F$ depending on the choice of $R$.

In general, we will write the operator $R$ in terms of an $r$-matrix taking values in $\mathfrak{f} \otimes \mathfrak{f}$
\begin{equation}
r = T_1 \wedge T_2 + T_3 \wedge T_4 + \ldots \ , \qquad \qquad T_r \wedge T_s = T_r \otimes T_s - T_s \otimes T_r \ ,
\end{equation}
where $T_a$ is a basis of $\mathfrak{f}$ and by the use of the wedge product we enforce the antisymmetry of the operator.
The operator $R$ is then defined using the bilinear form as
\begin{equation}
R X = \Tr_2(r (1 \otimes X)) \ ,
\end{equation}
where $\Tr_2$ denotes that the contraction is taken over the second space in the tensor product $\mathfrak{f} \otimes \mathfrak{f}$.

To find the non-abelian dual
of the symmetric space \sm \eqref{ssslag} with respect to a subgroup $H
\subset F$ we first write the group element $f$ as
\begin{equation}
f = h f' \ , \qquad \qquad h \in H \ , \qquad f' \in F \ .
\end{equation}
Substituting this into the Lagrangian \eqref{ssslag} and gauging the global $H$ symmetry,
i.e. replacing $h^{-1} d h \to A$, where $A \in \mathfrak{h} = Lie(H)$,
we find as in \rf{3}
\begin{equation}\label{nadlag}
\mathcal{L} = \Tr\big[(f'^{-1} A_+ f' + f'^{-1} \partial_+f')P(f'^{-1} A_- f' + f'^{-1} \partial_- f') + 
%v3
v F_{-+} (A)\big] \ .
\end{equation}
Here $v$ is a Lagrange multiplier imposing the flatness of the connection
$A$, i.e. 
%v3
$F_{-+}(A) \equiv \partial_- A_+ - \partial_+ A_- + [A_-,A_+]=0$, and hence the
equivalence of \eqref{nadlag} and \eqref{ssslag}.
In general, the algebra
$\mathfrak{h}$ need not be semi-simple. In this case $v$ should not be taken
in the algebra $\mathfrak{h}$, but rather in its dual $\bar{\mathfrak{h}}$.
Introducing $T_a$ as a basis for $\mathfrak{f}$ and $e_\a$ as a basis for $\mathfrak{h}$
we can define a basis of $\bar{\mathfrak{h}}$ to be
\begin{equation}\label{dualbasis}
\bar e_\a = \sum_{a=1}^{\dim F} \Tr[T_a e_\a] \, T_a \ ,
\end{equation}
as $F$ is assumed to be semi-simple.
The NAD model (defined on a flat 2d background, cf. \rf{3} and \rf{5}) is then found upon
integrating out the gauge field $A$
in \eqref{nadlag}.

The Lagrangian \eqref{nadlag} still has a local $G$ gauge symmetry $f' \to f'
g$, but now it also possesses an additional local $H$ gauge symmetry: $f' \to h f'$,
$A \to h A h^{-1} - dh h^{-1}$, $v \to h v h^{-1}$.\foot{One may be concerned
that the action of $H$ on $v$ does not preserve $v \in \bar{\mathfrak{h}}$ if
$\mathfrak{h} \neq \bar{\mathfrak{h}}$.
Indeed, this may be the case, but
one can always write the new terms as the sum of a part valued in
$\bar{\mathfrak{h}}$ and a part that drops out in the bilinear form. To show
this we first write an algebra element $X \in \mathfrak{f}$ as $X = X^\a \bar
e_\a + Y^{I} t_{I}$ where $t_{I}$ is some extension of $\bar e_\a$ to a basis
of $\mathfrak{f}$. Then if $\Tr[t_{I} e_{\a}] = c \neq 0$ for some $I$ and $\a$,
it follows from \eqref{dualbasis} that $\bar e_\a = c t_{I} + \ldots$. We can
then define a new generator $\tilde t_I = t_I - c \bar e_\a / \sum_{a=1}^{\dim
F} \Tr[T_a e_\a]^2$, which does satisfy $\Tr[\tilde t_I e_\a] = 0$. Replacing
$t_I \to \tilde t_I$ in the basis of $\mathfrak{f}$ still gives a basis of
$\mathfrak{f}$. Therefore, applying this process iteratively, we can construct a
basis $(\bar e_\a$, $\tilde t_I)$ with the desired property $\Tr[\tilde t_I
e_\a] = 0$.}
In the case when the NAD is with respect to the full group $H = F$, the $H$
gauge symmetry may be used to fix $f' = 1$. In this gauge fixing the $G$ gauge
symmetry then has a compensating action on $A$ and $v$ which may be used to
fix $v$.
When $H$ is a subgroup one may no
longer be able to fix $f' = 1$. The gauge condition may, in general, involve both $f'$ and $v$.
Our approach will be to gauge fix $f'$ as far as possible
and use the remaining gauge symmetry to constrain $v$.

As suggested by the example of the TsT transformation discussed in section \ref{sectst}, in
some cases it will not be enough to perform the NAD in a subgroup $H$ of $F$.
Rather we will need to start with a central extension of $H$ (which, in general, need not
be admissible in the full group).
We shall define the NAD with respect to the centrally-extended group $H_{c.e.}$ as (cf. \rf{13})\foot{One %v2
may attempt to promote the Lagrange multiplier terms to a gauged WZW model as in
\cite{sfet} in order to construct a $\lambda$-model type theory with an additional deformation parameter
based on \eqref{nadlagce}. It would be interesting to see if this preserves integrability, or if one also needs
to modify the first term. There may also be subtleties for centrally-extended and non-semi-simple
algebras.}
\begin{equation}\begin{split}\label{nadlagce}
{\hat {\mathcal{L} }}& = \Tr\big[(f'^{-1} A_+ f' + f'^{-1} \partial_+f')P(f'^{-1} A_- f' + f'^{-1} \partial_- f') +
%v3
 v F_{-+} (A)\big]
\\ & \qquad + w_m (\partial_- C^m_+ - \partial_+ C^m_-) + w_m [A_-,A_+]_{c.e.}^m \ .
\end{split}\end{equation}
Here the first line is identical to \eqref{nadlag}, i.e.
no central extension is present and
the Lie brackets remain those of $\mathfrak{f}$
and the subalgebra $\mathfrak{h}$.
In the second line the index $m$ labels the central extensions of $\mathfrak{h}$, while $w_m$ and $C^m$ are the
corresponding Lagrange multipliers and gauge fields respectively. The bracket
$[\ ,\, ]_{c.e.}$ is then the Lie bracket on the centrally-extended algebra $\mathfrak{h}_{c.e.}$.

On integrating out the Lagrange multipliers $v$ and $w_m$ in \eqref{nadlagce} we still
recover the symmetric space \sm \eqref{ssslag}, demonstrating that, like \rf{nadlag},
\rf{nadlagce} is still equivalent to \rf{ssslag}.
However,
integrating out the gauge fields in \eqref{nadlagce} will
now lead to a more general model (depending on extra parameters)
than that found from \eqref{nadlag}.
Indeed, following the discussion in section \ref{sectst}, integrating out $C^m$
implies that $w_m=w^{(0)}_m=\const$.
Substituting this back into \eqref{nadlagce} gives
\begin{equation}\label{nadfin}
{\hat {\mathcal{L} }} = \Tr\big[(f'^{-1} A_+ f' + f'^{-1} \partial_+f')P(f'^{-1} A_- f' + f'^{-1} \partial_- f') 
%v3
+ v F_{-+} (A)\big] + w^{(0)}_m [A_-,A_+]^m_{c.e.} \ .
\end{equation}
Then integrating out the gauge field $A_\pm$
defines the NAD model, which will now depend on arbitrary constant parameters $w_m^{(0)}$.
As was shown in section \ref{sectst}, in the case when $\mathfrak{h}$ is abelian these constants can
then be interpreted as the parameters of the TsT transformations.

%%%%%%%%%%%%%%%%%%%%%%%%%%%%%%%%%%%%%%
\subsection{\texpdf{$AdS_5$}{AdS5} and \texpdf{$\mathfrak{so}(2,4)$}{so(2,4)}}\label{seccon}
%%%%%%%%%%%%%%%%%%%%%%%%%%%%%%%%%%%%%%

Let us now turn to the $F/G= SO(2,4)/SO(1,4)$ symmetric space \sm for $AdS_5$ and first
briefly state our conventions for the algebra $\mathfrak{so}(2,4)$.
Introducing the $\g$-matrices (here $\s_{1,2,3}$ are the standard Pauli matrices and
$\s_0$ is the $2 \times 2$ identity matrix)
\begin{equation}
\g_0 = i \s_3 \otimes \sigma_0 \ , \quad \g_1 = \s_2 \otimes \s_2 \ , \quad \g_2 = - \s_2 \otimes \s_1 \ , \quad \g_3 = \s_1 \otimes \s_0 \ , \quad \g_4 = \s_2 \otimes \s_3 \ , %v3
\end{equation}
we define the following basis for $\mathfrak{so}(2,4)$
\begin{equation}
\textstyle T_{ij} = \frac14 [\g_i, \g_j] \ , \quad T_{i5} = -T_{5i} = \frac12 \g_i \ , \quad T_{55} = 0 \ , \qquad i,j = 0,\ldots 4 \ .
\end{equation}
We then use the standard matrix trace as our bilinear form.
It is this basis that we use to define bases for the duals of non semi-simple subalgebras of
$\mathfrak{so}(2,4)$ as in \eqref{dualbasis}.

The deformations based on non-abelian $r$-matrices that we consider are
more naturally understood in the Poincar\'e patch of $AdS_5$. Therefore, we will use
the corresponding basis of $\mathfrak{so}(2,4)$
\begin{equation}
D = T_{45} \ , \qquad P_\mu = T_{\mu 5} - T_{\mu 4} \ , \qquad K_\mu = T_{\mu 5} + T_{\mu 4} \ , \qquad M_{\mu \nu} = T_{\mu \nu} \ , \qquad \mu,\nu = 0,\ldots3 \ .
\end{equation}
The symmetric space $AdS_5$ can be represented as the coset $SO(2,4)/SO(1,4)$. The subalgebra
$\mathfrak{so}(1,4)$ corresponding to the gauge group is spanned by $T_{ij}$ where $i,j = 0,\ldots , 4$
and therefore the projector $P$ onto the coset part of the algebra is given by
\begin{equation}
P(X) = - \Tr[X T_{05}] T_{05} + \sum_{i=1}^4 \Tr[X T_{i5}] T_{i5} \ .
\end{equation}
Taking the gauge-fixed field $f$ to be
\begin{equation}\label{f}
f = \exp[- x_0 P_0 + x_1 P_1 + x_2 P_2 + x_3 P_3]\, \exp[\log z \, D] \ ,
\end{equation}
and substituting it into the Lagrangian \eqref{ssslag} of the symmetric space \sm we find that it takes the form of the \sm with the
target space metric being
the $AdS_5$ metric in Poincar\'e patch
\begin{equation}
ds^2 = \frac{- dx_0^2 + dx_1^2 + dx_2^2 + dx_3^2 + dz^2}{z^2} \ .
\end{equation}

%%%%%%%%%%%%%%%%%%%%%%%%%%%%%%%%%%%%%%
\subsection{Abelian \texpdf{$r$}{r}-matrices}\label{secabe}
%%%%%%%%%%%%%%%%%%%%%%%%%%%%%%%%%%%%%%

The first examples of the relation between homogeneous YB deformations and NAD
that we will consider are based on abelian $r$-matrices. While the equivalence in
these cases follows from the results of section \ref{sectst} and
\cite{Osten:2016dvf}, it will be instructive to consider two examples
explicitly. Prior to the general investigation of
\cite{Osten:2016dvf}, deformations based on
abelian $r$-matrices and their relation to TsT transformations of \adss have been studied
extensively on a case by case basis
\cite{yabe1,yabe2,ysum,vtsum,ysugra1,Hoare:2016hwh,ysugra2}.

{\bf Example 1:} \ The first case we shall consider corresponds to the rank 2 abelian $r$-matrix
\begin{equation}\la{321}
r = \eta \, P_2 \wedge P_3 \ ,
\end{equation}
where $P_\mu$ are translation generators and $\eta$ is a free deformation parameter. This $r$-matrix
and also the rank 4 one below \eqref{rmatabel4} were first discussed in \cite{yabe2} where it was shown
that the metrics and $B$-fields of corresponding YB deformed models are those
of the ``non-commutative dual''
backgrounds of \cite{Hashimoto:1999ut,Maldacena:1999mh}.
Using the gauge-fixed field $f$ \eqref{f}, the
Lagrangian of the corresponding YB deformed model \eqref{yblag} with $r$-matrix given by \rf{321}
is found to be
that of the \sm with the following target space metric and $B$-field (cf. \rf{8})
\begin{equation}\begin{split}\label{abel2}
ds^2 & = \frac{-dx_0^2 + dx_1^2 + dz^2}{z^2} + \frac{z^2}{z^4 +\eta^2} (dx_2^2 + dx_3^2) \ ,
\\
B & = \frac{\eta}{z^4 + \eta^2} dx_2 \wedge dx_3 \ .
\end{split}\end{equation}
On the other hand, let us consider the NAD of the $AdS_5$ \sm
with respect to the central
extension of the algebra $\mathfrak{h} = \{P_2,P_3\}$ (equivalent to \rf{9}). In this case
the dual algebra is given by $\bar{\mathfrak{h}} = \{K_2,K_3\}$.
We use the local $H$ gauge symmetry of the NAD model \eqref{nadfin} to fix
\begin{equation}
f' = \exp[-x_0 P_0 + x_1 P_1] \exp[\log z \, D] \ ,
\end{equation}
and also parametrize the gauge field and Lagrange multiplier as
\begin{equation}
A_\pm = A_{1\pm} P_2 + A_{2\pm} P_3 \ , \qquad \qquad v = \frac{1}{2\eta}(x_2 K_2 + x_3 K_3) \ .
\end{equation}
Substituting these expressions into the Lagrangian \eqref{nadfin} where we take the explicit form
of the central extension term to be (cf. \rf{13})
\begin{equation}\label{2ce}
w_m^{(0)} [A_-,A_+]_{c.e.}^m = \frac{1}{\eta}(A_{1+}A_{2-} - A_{2+}A_{1-}) \ ,
\end{equation}
and integrating out the gauge field we find that the NAD of $AdS_5$ with
respect to the central extension of $\mathfrak{h} = \{P_2,P_3\}$ again gives
the \sm based on the background metric and $B$-field \eqref{abel2} (up to a
total derivative term in the $B$-field).\foot{In what follows we will always ignore total derivative terms when comparing the
$B$-fields.}

{\bf Example 2:} \ Our second example is defined by the rank 4 abelian $r$-matrix
\begin{equation}\label{rmatabel4}
r = \eta \, P_0 \wedge P_1 + \zeta \, P_2 \wedge P_3 \ ,
\end{equation}
where $\eta$ and $\zeta$ are independent parameters.
Using again the gauge-fixed field $f$ \eqref{f},
the deformed metric and $B$-field corresponding to the YB deformed model \eqref{yblag}
are found to be
\begin{equation}\begin{split}\label{abel4}
ds^2 & = \frac{z^2}{z^4 - \eta^2} (- dx_0^2 + dx_1^2) + \frac{z^2}{z^4 + \zeta^2} ( dx_2^2 + dx_3^2) + \frac{dz^2}{z^2} \ ,
\\
B & = \frac{\eta}{z^4 - \eta^2} dx_0 \wedge dx_1 + \frac{\zeta}{z^4 + \zeta^2} dx_2 \wedge dx_3 \ .
\end{split}\end{equation}
Next, we construct the NAD of the $AdS_5$ \sm with respect to central extension of the
algebra $\mathfrak{h} = \{P_0, P_1; P_2,P_3 \}$ which is implied
by the form of the $r$-matrix \eqref{rmatabel4}
\begin{equation}\label{ceabel4}
[P_0,P_1] = Z_1 \ , \qquad\qquad [P_2,P_3] = Z_2 \ .
\end{equation}
The dual algebra is given by $\bar{\mathfrak{h}} = \{K_0,K_1;K_2,K_3\}$.
We now use the local $H$ gauge symmetry of \eqref{nadfin} to fix
\begin{equation}
f' = \exp[\log z \, D] \ ,
\end{equation}
and parametrize the gauge field and the Lagrange multiplier in \rf{nadfin}
as
\begin{equation}
A_\pm = A_{1\pm} P_0 + A_{2\pm} P_1 + A_{3\pm} P_2 + A_{4\pm} P_3 \ , \qquad v = \frac{1}{2\eta}(x_1 K_0 - x_0 K_1) + \frac{1}{2\zeta}(x_2 K_2 + x_3 K_3) \ .
\end{equation}
We then
substitute these expressions into the Lagrangian \eqref{nadfin} where, according to \eqref{ceabel4}
the explicit form of the central extension term is now
\begin{equation}\label{4ce}
w_m^{(0)} [A_-,A_+]_{c.e.}^m = \frac{1}{\eta}(A_{1+}A_{2-} - A_{2+}A_{1-}) + \frac{1}{\zeta}(A_{3+}A_{4-} - A_{4+}A_{3-}) \ .
\end{equation}
Integrating out the gauge field $A$ we conclude that the NAD transform of $AdS_5$ \sm with respect to
the central extension of $\mathfrak{h} = \{P_0,P_1;P_2,P_3\}$
gives the \sm based again on the metric
and $B$-field in \eqref{abel4}.
In addition, in both examples the NAD
procedure determines also the dilaton field given by the standard T-duality expression \cite{bu} (cf. \rf{5},\rf{8}).

%%%%%%%%%%%%%%%%%%%%%%%%%%%%%%%%%%%%%%
\subsection{Unimodular non-abelian \texpdf{$r$}{r}-matrices}\label{secuni}
%%%%%%%%%%%%%%%%%%%%%%%%%%%%%%%%%%%%%%

Let us now turn our attention to homogeneous YB deformations based on
non-abelian $r$-matrices. We start with unimodular examples \eqref{uni} for
which the YB deformation of the supercoset \sm background \cite{yjor,dmv}
preserves the satisfaction of the supergravity equations or one-loop Weyl
invariance (with an appropriate dilaton) \cite{Borsato:2016ose}.

All abelian $r$-matrices, defining deformations which are equivalent to sequences of TsT transformations of \adss
\cite{Osten:2016dvf}, are unimodular. Furthermore, at rank 2 all unimodular
$r$-matrices are abelian. However, for the algebra $\mathfrak{so}(2,4)$ there
are rank 4 and rank 6 non-abelian unimodular $r$-matrices.
Those of rank 4 were classified in \cite{Borsato:2016ose} and fall into three
classes characterized by the algebra of the corresponding
generators. Defining the $r$-matrix
as
\begin{equation}
r = e_1 \wedge e_2 + e_3 \wedge e_4 \ ,
\end{equation}
the non-vanishing commutation relations determining the three classes are\foot{In \cite{Borsato:2016ose}
the $r$-matrices are grouped into four classes, with the non-vanishing commutation relations
$[e_1, e_3] = e_4$, $[e_1, e_4] = e_3$ determining the additional class. Here we note that
these commutation relations are related to those of class 2 by analytic continuation and hence for
our purposes do not need to be considered independently.\label{footcomrel}}
\begin{align} \nonumber
& \text{ class 1:} \quad [e_1, e_4] = e_2 \ ,
\\
& \text{ class 2:} \quad  [e_1, e_4] = e_3  \ , \qquad  [e_1, e_3] = - e_4 \ , \la{333}
\\ \nonumber
& \text{ class 3:} \quad [e_1, e_4] = e_2 \ , \qquad [e_1, e_3] = -e_4 \ .
\end{align}
In \cite{Borsato:2016ose} the deformations
corresponding to examples from the first two classes were observed to be equivalent to a sequence of two TsT
transformations (with a non-linear coordinate
redefinition in between) of the \adss model; however,
a similar result for the last class was not found.

Let us now consider one example from each class
demonstrating that the corresponding YB deformation is equivalent to the NAD of the $AdS_5$ \sm with respect
to the following central extension of the algebra $\mathfrak{h} = \{e_1,e_2;e_3,e_4\}$\foot{One can check that the
Jacobi identity is satisfied and this central extension is consistent for all three
classes.}
\begin{equation}
[e_1,e_2] = Z_1 \ , \qquad \qquad [e_3,e_4] = Z_2 \ .
\end{equation}
As mentioned in section \ref{sectst}, centrally extending a unimodular algebra
preserves this property. At the end of this section we will also consider one
rank 6 example.

{\bf Class 1 example:} \ An $r$-matrix from class 1 that we shall consider is \cite{Borsato:2016ose}
\begin{equation}\label{class1rmat}
r = \eta \, M_{+3} \wedge P_+ + \zeta \, P_2 \wedge P_3 \ ,
\end{equation}
where $\eta$ and $\zeta$ are two free parameters.
Here we have introduced light-cone indices in the $\mathfrak{so}(1,3)$ Lorentz subalgebra spanned by $M_{\mu \nu}$
(not be confused with the light-cone coordinates on the worldsheet) defined as
$\Lambda_\pm = \Lambda_0 \pm \Lambda_1 $,
i.e. $M_{+3}= M_{03} + M_{13}$.
Fixing the group-valued field $f$ as
\begin{equation}
f = \exp[-\ha(x_- P_+ + x_+ P_-) + x_2 P_2 + x_3 P_3] \exp[\log z \, D] \ ,
\end{equation}
we find the YB deformed model \eqref{yblag} corresponds to following metric and $B$-field
\begin{equation}\begin{split}\label{class1back}
ds^2 & = \frac{-dx_- dx_+ + dz^2}{z^2} + \frac{- \eta^2 x_+^2 dx_+^2 - 2 \eta \zeta x_+ dx_+ dx_2 + z^4(dx_2^2 + dx_3^2) }{z^2(z^4 + \zeta^2)} \ ,
\\
B & = \frac{\eta x_+ dx_+ \wedge dx_3 + \zeta dx_2 \wedge dx_3} {z^4 +\z^2} \ .
\end{split}\end{equation}
Let us now compare this to the NAD of the $AdS_5$ \sm with respect to the central extension of
the algebra $\mathfrak{h} = \{M_{+3}, P_+; P_2, P_3\}$.
The dual algebra is given by
$\bar{\mathfrak{h}} = \{ M_{-3}, K_-; K_2, K_3\}$. We first partially use the local
$H$ gauge symmetry of \eqref{nadfin} to fix $f'$ to be
\begin{equation}
f' = \exp[-\ha x_+ P_-] \exp[\log z \, D] \ .
\end{equation}
This leaves one gauge freedom corresponding to $M_{+3}$ that we cannot use to
fix $f'$ further. We therefore use it to fix the $K_3$ component of the
Lagrange multiplier $v$ to zero, so that it can be parametrized as
\begin{equation}\la{439}
v = -\big(\frac{x_-}{4\eta} + \frac{x_+x_2}{2\zeta}\big) M_{-3} + \frac12\sqrt{\frac{-x_2}{2\eta\zeta}} \, K_-
+ \big(\frac{x_3}{2\zeta} + \frac{\eta x_+}{\zeta} \sqrt{\frac{-x_2}{2\eta\zeta}}\big) K_2 \ .
\end{equation}
The reason for this choice is to make manifest the comparison with \rf{class1back}.
Parametrizing the gauge field as
\begin{equation}\label{aa1}
A_\pm = A_{1\pm} M_{+3} + A_{2\pm} P_+ + A_{3\pm} P_2 + A_{4\pm} P_3 \ ,
\end{equation}
we again take the explicit form of the central extension term in \eqref{nadfin}
to be given by \eqref{4ce}.
Integrating out the gauge field we finally find that the
NAD of the $AdS_5$ \sm with respect to the central extension of $\mathfrak{h} =
\{M_{+3},P_+;P_2,P_3\}$ gives exactly the same \sm as defined by the
metric and $B$-field in \eqref{class1back}.

{\bf Class 2 example:} \ An example of $r$-matrix from class 2 is \cite{Borsato:2016ose}
\begin{equation}\label{class2rmat}
r = \eta \, M_{23} \wedge P_1 + \zeta \, P_2 \wedge P_3 \ .
\end{equation}
Fixing the group-valued field $f$ as in \rf{f} with
\begin{equation}\la{341}
x_2 = r \cos \theta \ , \qquad x_3 = r \sin \theta \ ,
\end{equation}
we find the following metric and $B$-field of the YB deformed model \eqref{yblag}
\begin{equation}\begin{split}\label{class2back}
ds^2 & = \frac{- dx_0^2 + dz^2}{z^2} + \frac{(z^4 + \z^2) dx_1^2 - 2 \eta \zeta r dx_1 dr + (z^4 + \eta^2 r^2) dr^2 + z^4 r^2 d \theta^2}{z^2(z^4 + \eta^2 r^2 + \zeta^2)} \ ,
\\
B & = \frac{\eta r^2 dx_1 \wedge d\theta + \zeta r dr \wedge d\theta} {z^4 + \eta^2 r^2 + \z^2} \ .
\end{split}\end{equation}
Next,
we compare this to the NAD of $AdS_5$ with respect to the central extension of
the algebra $\mathfrak{h} = \{M_{23}, P_1; P_2, P_3\}$. The dual algebra is given by
$\bar{\mathfrak{h}} = \{ M_{23}, K_1; K_2, K_3\}$. We partially use the local
$H$ gauge symmetry of \eqref{nadfin} to fix
\begin{equation}
f' = \exp[-x_0 P_0] \exp[\log z \, D] \ .
\end{equation}
This leaves one gauge freedom corresponding to $M_{23}$ that we
can use to set the $K_3$ component of the
Lagrange multiplier $v$ to zero, which we then parametrize as
\begin{equation}
v = \big(-\frac{x_1}{\eta} + \frac{r^2}{2\zeta}\big) M_{23} + \frac{\theta}{2\eta} K_1 + \frac{r}{2\zeta} K_2 \ .
\end{equation}
Parametrizing the gauge field as
\begin{equation}\label{aa2}
A_\pm = A_{1\pm} M_{23} + A_{2\pm} P_1 + A_{3\pm} P_2 + A_{4\pm} P_3 \ ,
\end{equation}
we again take the explicit form of the central extension term in \eqref{nadfin}
to be given by \eqref{4ce}. Integrating out the gauge field we find that the
NAD of the $AdS_5$ \sm with respect to the central extension of $\mathfrak{h} =
\{M_{23},P_1;P_2,P_3\}$ gives the \sm defined by the 
metric and $B$-field \eqref{class2back}.

Let us note that one can also consider the $r$-matrix
\begin{equation}\label{class2rmatalt}
r = \eta \, M_{01} \wedge P_3 + \zeta \, P_0 \wedge P_1 \ ,
\end{equation}
which can be understood as an analytic continuation of \eqref{class2rmat}. The corresponding
quasi-Frobenius algebra satisfies the commutation relations given in footnote \ref{footcomrel}.
By repeating the above discussion using the obvious
analytic continuation it is clear that the corresponding
YB deformed model is equivalent to the NAD of $AdS_5$ with respect to the central
extension of $\mathfrak{h} = \{M_{01}, P_3 ; P_0, P_1\}$.

{\bf Class 3 example:} \ Let us now consider the following
$r$-matrix from class 3\foot{This is the fourth example in \cite{Borsato:2016ose}
for which a TsT interpretation has thus far not been found.}
\begin{equation}\label{class3rmat}
r = \eta \, M_{+3} \wedge P_+ + \zeta \, P_1 \wedge P_3 \ .
\end{equation}
Fixing the group-valued field $f$ as in \rf{f}, i.e.
\begin{equation}
f = \exp[-\ha(x_- P_+ + x_+ P_-) + x_2 P_2 + x_3 P_3 ] \exp[\log z \, D] \ ,
\end{equation}
we find the following metric and $B$-field of the YB deformed model \eqref{yblag}
\begin{align}\nonumber
ds^2 & = \frac{dx_2^2 + dz^2}{z^2} + \frac{-2(2z^4 +2\eta\zeta x_+ + \zeta^2)dx_-dx_+ -\zeta^2 dx_-^2 - (2\eta x_+ + \zeta)^2 dx_+^2 + 4z^4 dx_3^2}{4z^2(z^4 + 2\eta \zeta x_+ + \zeta^2)} \ ,
\\\label{class3back}
B & = \frac{-\zeta dx_- \wedge dx_3 + (2\eta x_+ + \zeta)dx_+ \wedge dx_3} {2(z^4 + 2\eta \zeta x_+ + \zeta^2)} \ .
\end{align}
We are now to compare this to the NAD of $AdS_5$ with respect to the central extension of
$\mathfrak{h} = \{M_{+3}, P_+; P_1, P_3\}$ with the dual algebra being
$\bar{\mathfrak{h}} = \{ M_{-3}, K_-; K_1, K_3\}$. We partially use the local
$H$ gauge symmetry of \eqref{nadfin} to fix
\begin{equation}
f' = \exp[x_2 P_2] \exp[\log z \, D] \ .
\end{equation}
This leaves one gauge freedom corresponding to $M_{+3}$ that we
can use this to fix the $K_3$ component of the
Lagrange multiplier $v$ to zero, which we then parametrize as
\begin{equation}
v = -\big(\frac{x_- + x_+}{4\eta} + \frac{x_+^2}{4\zeta}\big) M_{-3} + \big(\frac{x_3}{2\zeta} + \sqrt{\frac{-x_+}{2\eta \zeta}} (1 + \frac{2 \eta x_+}{3\zeta}) \big) K_-
+ \big(\frac{x_3}{\zeta} + \sqrt{\frac{-x_+}{2\eta\zeta}} (1+ \frac{4\eta x_+}{3\zeta}) \big) K_1 \ .
\end{equation}
Parametrizing the gauge field as
\begin{equation}\label{aa4}
A_\pm = A_{1\pm} M_{+3} + A_{2\pm} P_+ + A_{3\pm} P_1 + A_{4\pm} P_3 \ ,
\end{equation}
choosing the central extension term in \eqref{nadfin}
to be \eqref{4ce} and integrating out the gauge field we finally
conclude that the
NAD of the $AdS_5$ \sm with respect to the central extension of $\mathfrak{h} =
\{M_{+3},P_+,P_1,P_3\}$ is again equivalent to the YB deformed \sm corresponding to the
metric and $B$-field in \eqref{class3back}.

{\bf Rank 6 example:} \ Finally, we consider the following example of a rank 6
unimodular $r$-matrix \cite{Borsato:2016ose}
\begin{equation}\label{rank6rmat}
r = \eta \, M_{01} \wedge M_{23} + \zeta \, P_0 \wedge P_1 + \kappa \, P_2 \wedge P_3 \ ,
\end{equation}
where $\eta$, $\zeta$ and $\kappa$ are free parameters.
Fixing the group-valued field $f$ as in \rf{f} with
\begin{equation}\la{341a}
x_0 = t \cosh \chi \ , \qquad x_1 = t \sinh \chi \ , \qquad x_2 = r \cos \theta \ , \qquad x_3 = r \sin \theta \ ,
\end{equation}
we find the following metric and $B$-field of the YB deformed model \eqref{yblag}
\begin{equation}\begin{split}\label{rank6back}
ds^2 & = \frac{ dz^2}{z^2} + \frac{-z^2(z^4 + \eta^2 t^2 r^2 + \kappa^2) dt^2 + 2 \eta \zeta z^2 t r^2 dt d\theta + (z^4- \zeta^2)z^2 r^2 d\theta^2 }{z^8 + z^4(\eta^2 t^2 r^2 - \zeta^2 + \kappa^2) - \zeta^2 \kappa^2}
\\ & \qquad \qquad
+ \frac{z^2(z^4 + \eta^2 t^2 r^2 - \zeta^2) dr^2 - 2 \eta \kappa z^2 t^2 r dr d\chi + (z^4+\kappa^2)z^2 t^2 d\chi^2 }{z^8 + z^4(\eta^2 t^2 r^2 - \zeta^2 + \kappa^2) - \zeta^2 \kappa^2} \ ,
\\
B & = \frac{-\eta \zeta \kappa t r dt \wedge dr + \zeta (z^4 + \kappa^2) t dt \wedge d\chi - \kappa (z^4 - \zeta^2) r d\theta \wedge dr - \eta z^4 t^2 r^2 d\theta \wedge d\chi } {z^8 + z^4(\eta^2 t^2 r^2 - \zeta^2 + \kappa^2) - \zeta^2 \kappa^2} \ .
\end{split}\end{equation}
Now we compare this to the NAD of
$AdS_5$ with respect to the central extension of
the algebra $\mathfrak{h} = \{M_{01},M_{23}; P_0, P_1; P_2, P_3\}$.\foot{The
central extension is given by $[M_{01},M_{23}] = Z_1$, $[P_0,P_1] = Z_2$, $[P_2,P_3] = Z_3$
as follows from the structure of the $r$-matrix \eqref{rank6rmat}. Again, one can check that the
Jacobi identity is satisfied.} The dual algebra is given by
$\bar{\mathfrak{h}} = \{M_{01}, M_{23}; K_0, K_1; K_2, K_3\}$. We partially use the local
$H$ gauge symmetry of \eqref{nadfin} to fix
\begin{equation}
f' = \exp[\log z \, D] \ .
\end{equation}
This leaves two free gauge transformations corresponding to $M_{01}$ and $M_{23}$ that we
can use to set the $K_0$\foot{Writing $v = v_0 K_0 + v_1 K_1 + \ldots$,
this is a valid gauge fixing in the ``patch'' in which $v_1^2 > v_0^2$.} and $K_2$ components of the
Lagrange multiplier $v$ to zero, which we then parametrize as
\begin{equation}
v = -\big(\frac{\theta}{\eta} - \frac{t^2}{2\zeta}\big) M_{01} - \big(\frac{\chi}{\eta} - \frac{r^2}{2\kappa}\big) M_{23}
+ \frac{t}{2\zeta} K_1 + \frac{r}{2\kappa} K_3 \ .
\end{equation}
Parametrizing the gauge field as
\begin{equation}\label{aarank6}
A_\pm = A_{1\pm} M_{01} + A_{2\pm} M_{23} + A_{3\pm} P_0 + A_{4\pm} P_1 + A_{5\pm} P_2 + A_{6\pm} P_3 \ ,
\end{equation}
we again take the explicit form of the central extension term in \eqref{nadfin}
to be given by
\begin{equation}\label{6ce}
w_m^{(0)} [A_-,A_+]_{c.e.}^m = \frac{1}{\eta}(A_{1+}A_{2-} - A_{2+}A_{1-}) + \frac{1}{\zeta}(A_{3+}A_{4-} - A_{4+}A_{3-})
+\frac{1}{\kappa}(A_{5+}A_{6-} - A_{6+}A_{5-}) \ .
\end{equation}
Integrating out the gauge field we find that the
NAD of the $AdS_5$ \sm with respect to the central extension of $\mathfrak{h} =
\{M_{01},M_{23};P_0,P_1;P_2,P_3\}$ gives the \sm defined by
metric and $B$-field \eqref{rank6back}.

%%%%%%%%%%%%%%%%%%%%%%%%%%%%%%%%%%%%%%
\subsection{Jordanian \texpdf{$r$}{r}-matrices}\label{secjor}
%%%%%%%%%%%%%%%%%%%%%%%%%%%%%%%%%%%%%%

We now turn to our final group of examples, the jordanian $r$-matrices. Jordanian
$r$-matrices have the form (see, e.g., \cite{jord} and references therein) %v2
\begin{equation}
r = T_1 \wedge T_2 + \ldots \ ,\qquad \qquad [T_1,T_2] = T_2 \ ,
\end{equation}
and the corresponding deformations of $AdS_5$ have been extensively
studied on a case by case basis in
\cite{yjor,ysum,vtsum,ysugra1,Hoare:2016hwh,ysugra2}.
When built out of
bosonic generators the jordanian $r$-matrices do not satisfy the
unimodularity property \eqref{uni} \cite{Borsato:2016ose}.
Indeed, the
backgrounds corresponding to the jordanian deformations of the supercoset
model \cite{yjor,dmv} do not solve the supergravity equations
\cite{ysugra1,Hoare:2016hwh,ysugra2}, solving instead the generalized equations
of \cite{Arutyunov:2015mqj,Wulff:2016tju}.

%v2
In the following we will compare the YB deformations arising from jordanian
$r$-matrices to the NAD of $AdS_5$ \sm with respect to the corresponding
quasi-Frobenius subalgebra itself (i.e. without central extension). A possible
reason why we do not need to consider central extensions is that, unlike for
the unimodular $r$-matrices, the central extension of interest turns out to be
trivial. Indeed, let us consider the simplest case of the $r$-matrix $D\wedge
P_0$.  The two generators here  have the commutation relation $[D,P_0] = P_0$.
If we try to centrally extend this 2d algebra we get $[D,P_0] = P_0 + Z$, but
now defining $P'_0 = P_0 + Z$ we see that this extension is trivial. For the
extended $r$-matrix $D \wedge P_0 + M_{01} \wedge P_1 + M_{+2} \wedge P_2 +
M_{+3} \wedge P_3$ we consider the central extension $[D,P_0] = P_0 + Z$,
$[M_{01},P_1] = P_0 + Z$, $[M_{+2},P_2] = P_+ + Z$ and $[M_{+3},P_3] = P_+ +
Z$. Here we only consider a single extension as there is only a single free
parameter scaling the whole $r$-matrix.  Again by shifting $P_0$ we see that
this extension is trivial. Similar statements hold for the remaining examples
that we consider.

As for the abelian $r$-matrices and unimodular non-abelian $r$-matrices
in sections \ref{secabe} and \ref{secuni} we will again provide exhaustive
evidence for the equivalence of the YB and NAD constructions.
In this section we will discuss two special cases
in detail, while in Appendix \ref{appjor} we will summarize the key information for twenty
additional examples.

{\bf Example 1:} \ The first example is the case of the rank 2 jordanian $r$-matrix
\begin{equation}\label{jorp0rmat}
r = \eta \, D \wedge P_0 \ ,
\end{equation}
where $D$ is the dilatation operator.
Parametrizing the group-valued field $f$ as in \rf{f} with
\begin{equation}\la{355}
x_1 = r \cos \theta \ , \qquad x_2 = r \sin\theta \, \cos \phi \ , \qquad x_3 = r \sin\theta \, \sin \phi \ ,
\end{equation}
we find the following metric and $B$-field of the corresponding YB deformed model \eqref{yblag}
\begin{equation}\begin{split}\label{jorp0back}
ds^2 & = \frac{-z^4 dx_0^2 + z^2(z^2 - \eta^2) dr^2 + 2 \eta^2 z r dz dr + (z^4-\eta^2 r^2) dz^2}{z^2(z^4 -\eta^2 z^2 - \eta^2 r^2)} + \frac{r^2}{z^2}(d\theta^2 + \sin^2\theta \, d\phi^2) \ ,
\\
B & = \frac{\eta r dr \wedge dx_0 + \eta z dz \wedge dx_0} {z^4 - \eta^2 z^2 - \eta^2 r^2} \ .
\end{split}\end{equation}
In \cite{ysugra2} it was observed that this background can be also obtained from the $AdS_5$ \sm
via a generalized ``TsT'' transformation in which the shift is replaced with a
more complicated non-linear field redefinition.\foot{The coordinate
transformation is required is to make the dilatation symmetry a linear
isometry.}

It turns out that this background can be derived in one step
by applying the NAD transformation to the $AdS_5$ \sm with respect to
the non-semisimple subalgebra $\mathfrak{h} = \{D, P_0\}$
(with the dual algebra being given by
$\bar{\mathfrak{h}} = \{D,K_0\}$). We first use the local
$H$ gauge symmetry of \eqref{nadlag} to fix $f'$ as
\begin{equation}\label{para1}
f'= \exp\big[\frac{1}{z}(x_1 \, P_1 + x_2 \, P_2 + x_3 \, P_3) \big]\ ,
\end{equation}
where $x_i$ are again given by \rf{355}. We also parametrize the Lagrange multiplier $v$ in terms of the
coordinates $x_0$ and $z$ as
\begin{equation}\label{rr0}
v=\eta^{-1} \big[- x_0 D + \ha {z} K_0\big] \ .
\end{equation}
Substituting this into \eqref{nadlag} and integrating out the gauge field, we
find that the NAD of the $AdS_5$ \sm with respect to $\mathfrak{h} = \{D,P_0\}$
gives the \sm based on the metric and $B$-field
 \eqref{jorp0back}.\foot{If we instead consider %v2
the centrally-extended algebra $[D,P_0] = P_0 + Z$,
which, as discussed  above, amounts to a trivial extension, (with $A_\pm = A_{1\pm}
D + A_{2\pm} P_0$ and the explicit form of the central extension term as
in \eqref{2ce}) we  again recover the background \eqref{jorp0back} after taking $f'$
as in \eqref{para1} and $v = \eta^{-1} [-x_0 D + \ha (z-1) K_0] $. This simple
shift in the Lagrange multiplier is directly correlated to the shift in $P_0$
required to reach the trivially-extended algebra. Similar statements should
hold for the remaining examples.}

The fact that the algebra $\mathfrak{h}$ is not unimodular (i.e. $n_a$ in \rf{4} is non-zero)
and hence the NAD transformation does not preserve Weyl invariance
is then recognized as the reason why
the full YB deformed \adss background found in \cite{ysugra2} solves only the generalized (scale-invariance)
but not standard supergravity (Weyl-invariance) equations.

From the analysis in section \ref{ssecnad} we expect that if we T-dualize the
full YB deformed \adss background in the direction of $v$ corresponding to
$n_\a = f^{\gamma}_{\ \gamma \alpha}$ we find a supergravity solution with a
linear dilaton in that direction. From the Lagrange multiplier \eqref{rr0} this
implies that the dilaton should be linear in $x_0$ (i.e. the coordinate
associated to the dilatation generator $D$). This is in agreement with the
results of \cite{ysugra2}.

{\bf Example 2:} \ As our final examples
let us consider the following pair of extended jordanian $r$-matrices \cite{Hoare:2016hwh}
\begin{align}
\label{jorext1rmat}
r & = \eta \, \big(D \wedge P_0 + M_{01} \wedge P_1 + M_{+2} \wedge P_2 + M_{+3} \wedge P_3\big) \ ,
\\ \label{jorext2rmat}
r & = \eta \, \big(D \wedge P_1 + M_{01} \wedge P_0 + M_{+2} \wedge P_2 + M_{+3} \wedge P_3\big) \ ,
\end{align}
which are closely related by an analytic continuation.
In both cases we fix the group-valued field $f$ as in \rf{f} with $x_2$ and $x_3$ written in terms of polar coordinates as in \rf{341}, i.e.
\begin{equation}
f= \exp[-x_0 P_0 + x_1 P_1 + r (\cos \theta \, P_2 + \sin \theta \, P_3)] \exp[\log z \, D] \ .
\end{equation}
Then the metrics and $B$-fields of the YB deformed model \eqref{yblag} are found to be
\begin{equation}\begin{split}\label{jorext1back}
ds^2 & = \frac{-dx_0^2 + dz^2}{z^2 - \eta^2} + \frac{z^2(dx_1^2 + dr^2)}{z^4 + \eta^2 r^2} + \frac{r^2}{z^2}d\theta^2 \ , \qquad
B = -\frac{\eta dx_0 \wedge dz}{z(z^2 - \eta^2)} + \frac{\eta r dx_1 \wedge dr}{z^4 + \eta^2 r^2} \ ,
\end{split}\end{equation}
and
\begin{equation}\begin{split}\label{jorext2back}
ds^2 & = \frac{dx_1^2 + dz^2}{z^2 + \eta^2} + \frac{z^2( - dx_0^2 + dr^2)}{z^4 - \eta^2 r^2} + \frac{r^2}{z^2}d\theta^2 \ , \qquad
B = - \frac{\eta dx_1 \wedge dz}{z(z^2 + \eta^2)} + \frac{\eta r dx_0 \wedge dr}{z^4 - \eta^2 r^2} \ ,
\end{split}\end{equation}
respectively.

These two $r$-matrices \eqref{jorext1rmat} and \eqref{jorext2rmat} are built from
the generators of the same algebra
\be \mathfrak{h} = \{D,M_{01},M_{+2},M_{+3},P_0,P_1,P_2,P_3\} \ , \ee
with the dual algebra being $\bar{\mathfrak{h}} = \{D,M_{01},M_{-2},M_{-3},K_0,K_1,K_2,K_3\} .$
Let us then consider the NAD of the $AdS_5$ \sm with respect to $\mathfrak{h}$.
After partially using the local $H$ gauge symmetry to completely fix $f'$ in \rf{nadlag}, i.e.
\begin{equation}
f' = 1 \ ,
\end{equation}
there are still three remaining gauge symmetries, corresponding to the generators $M_{01}$, $M_{+2}$, $M_{+3}$.
The Lagrange multiplier $v \in \bar{\mathfrak{h}}$ contains a piece $v_0 K_0 + v_1 K_1 + \ldots$.
For $v_0^2 > v_1^2$ we can use the remaining gauge symmetries to fix
\begin{equation}\label{regime1}
v =\eta^{-1} \big[ - x_0 D + \ha {z} K_0 + {x_1} M_{01} +\ha r (\cos \theta \, M_{-2} + \sin \theta \, M_{-3})\big] \ ,
\end{equation}
while for $v_1^2 > v_0^2$ we fix
\begin{equation}\label{regime2}
v =\eta^{-1} \big[ {x_1} D + \ha {z} K_1 - x_0 M_{01} +\ha r (\cos \theta \, M_{-2} + \sin \theta \, M_{-3})\big] \ .
\end{equation}
Substituting $v$ in \eqref{regime1} into \eqref{nadlag} and
integrating out the gauge field we find the \sm based on the
metric and $B$-field \eqref{jorext1back}, while using $v$
in \eqref{regime2} we recover the
\sm based on \eqref{jorext2back}.
%We conclude that
%v3
Thus the two YB deformations correspond to different ``patches''
of the NAD of $AdS_5$ model with respect to $\mathfrak{h}$.

To conclude, let us observe that for the jordanian examples the parameter
$\eta$, which appears in the $r$-matrix as an overall factor, also enters the
dual model in a similar way, with the Lagrange multiplier $v \sim \eta^{-1}$ (cf.
\eqref{rr0}, \eqref{regime1} and \eqref{regime2}). This is consistent as using the
automorphism $P_\m \to \lambda P_\m$, $K_\m \to \lambda^{-1} K_\m$ of $\mathfrak{so}(2,4)$
we can set $\eta = 1$ in the $r$-matrix, while the same can be done in the NAD
model by rescaling $v$. Similar observations also hold for the examples in
Appendix \ref{appjor} where in some cases we also use the inner automorphism generated by
the $SO(1,1)$ subgroup of the $SO(1,3)$ Lorentz algebra.

%%%%%%%%%%%%%%%%%%%%%%%%%%%%%%%%%%%%%%
\section*{Acknowledgments}
%%%%%%%%%%%%%%%%%%%%%%%%%%%%%%%%%%%%%%

We would like to thank R. Borsato, S. van Tongeren and L. Wulff for discussions of related questions and useful comments on the draft.
We also thank the organizers of the conference and focus program on ``Integrability in Gauge and String Theory''
at Humboldt University Berlin 23/08-02/09/2016 for kind hospitality.
The work of BH is partially supported by grant no. 615203 from the European Research Council under the FP7.
The work of AAT was supported by the ERC Advanced grant no. 290456, the STFC Consolidated grant ST/L00044X/1 and by the Russian Science Foundation grant 14-42-00047.

%%%%%%%%%%%%%%%%%%%%%%%%%%%%%%%%%%%%%%
\appendix
%%%%%%%%%%%%%%%%%%%%%%%%%%%%%%%%%%%%%%

%%%%%%%%%%%%%%%%%%%%%%%%%%%%%%%%%%%%%%
\section{Further examples corresponding to jordanian \texpdf{$r$}{r}-matrices}\label{appjor}
\def\theequation{A.\arabic{equation}}
\setcounter{equation}{0}
%%%%%%%%%%%%%%%%%%%%%%%%%%%%%%%%%%%%%%

In this Appendix we present a number of further examples of the relation
between YB deformed models based on jordanian $r$-matrices and non-abelian
duals of the $AdS_5$ $\s$-model. This list (including the cases discussed already
in section \ref{secjor}), while not a classification, covers the majority of the $r$-matrices
considered in \cite{yjor,ysum,vtsum,ysugra1,Hoare:2016hwh,ysugra2} up to
certain automorphisms of $\mathfrak{so}(2,4)$ including those based on $P_\m
\leftrightarrow K_\m$ and $+ \leftrightarrow -$.

For each case we will give the $r$-matrix and the
parametrization of the group element $f$ for the YB deformed model
\eqref{yblag}. We will then provide the corresponding data for the NAD model
\eqref{nadlag}: the algebra $\mathfrak{h}$ in which we dualize, its dual
$\bar{\mathfrak{h}}$ and the parametrizations of the group element $f'$ and of
the Lagrange multiplier $v \in \bar{\mathfrak{h}}$. As was mentioned in section
\ref{secjor}, for the jordanian $r$-matrices we do not need to consider central
extensions of $\mathfrak{h}$.

For reasons of brevity we will not present the explicit forms of the metrics and
$B$-fields, but just state that in all cases
one finds the complete agreement between the YB deformed
and NAD transformed $AdS_5$ model.

Note that the pair of examples 1 and 2 below follow a similar pattern as
the extended jordanian $r$-matrices in example 2 in section \ref{secjor}:
the two YB deformed models both correspond to the same NAD model.
The difference on the NAD side appears in the gauge fixing of
the Lagrange multiplier $v$.
The same is true also for the pair of examples 3 and 4.

\footnotesize

\begin{changemargin}{-0.57cm}{-1.43cm}
\begin{flalign*}
1. \quad \vphantom{\frac{z}{\eta}}
r & = \eta \, ( D \wedge P_0 + M_{01} \wedge P_1 + M_{+2} \wedge P_2) \ , &
f & = \exp[-x_0 P_0 + x_1 P_1 + x_2 P_2 + x_3 P_3]\exp[\log z \, D] \ , &
\\ \vphantom{\frac{z}{\eta}}
\mathfrak{h} & = \{D, P_0, M_{01}, P_1, M_{+2}, P_2\} \ , &
f' & = \exp[\frac{x_3}{z} P_3] \ , &
\\ \vphantom{\frac{z}{\eta}}
\bar{\mathfrak{h}} & = \{D, K_0, M_{01}, K_1, M_{-2}, K_2\} \ , &
v & = -\frac{x_0}\eta D + \frac{z}{2\eta} K_0 + \frac{x_1}{\eta} M_{01} + \frac{x_2}{2\eta} M_{-2} \ . &
\end{flalign*}

\begin{flalign*}
2. \quad \vphantom{\frac{z}{\eta}}
r & = \eta \, ( D \wedge P_1 + M_{01} \wedge P_0 + M_{+2} \wedge P_2) \ , &
f & = \exp[-x_0 P_0 + x_1 P_1 + x_2 P_2 + x_3 P_3]\exp[\log z \, D] \ , &
\\ \vphantom{\frac{z}{\eta}}
\mathfrak{h} & = \{D, P_1, M_{01}, P_0, M_{+2}, P_2\} \ , &
f' & = \exp[\frac{x_3}{z} P_3] \ , &
\\ \vphantom{\frac{z}{\eta}}
\bar{\mathfrak{h}} & = \{D, K_1, M_{01}, K_0, M_{-2}, K_2\} \ , &
v & = \frac{x_1}\eta D + \frac{z}{2\eta} K_1 - \frac{x_0}{\eta} M_{01} + \frac{x_2}{2\eta} M_{-2} \ . &
\end{flalign*}

\begin{flalign*}
3. \quad \vphantom{\frac{z}{\eta}}
r & = \eta \, ( D \wedge P_0 + M_{01} \wedge P_1) \ , &
f & = \exp[-x_0 P_0 + x_1 P_1 + r (\cos \theta \, P_2 + \sin \theta \, P_3)]\exp[\log z \, D] \ , &
\\ \vphantom{\frac{z}{\eta}}
\mathfrak{h} & = \{D, P_0, M_{01}, P_1\} \ , &
f' & = \exp[\frac{r}{z}(\cos \theta \, P_2 + \sin \theta \, P_3)] \ , &
\\ \vphantom{\frac{z}{\eta}}
\bar{\mathfrak{h}} & = \{D, K_0, M_{01}, K_1\} \ , &
v & = - \frac{x_0}\eta D + \frac{z}{2\eta} K_0 + \frac{x_1}{\eta} M_{01} \ . &
\end{flalign*}

\begin{flalign*}
4. \quad \vphantom{\frac{z}{\eta}}
r & = \eta \, ( D \wedge P_1 + M_{01} \wedge P_0) \ , &
f & = \exp[-x_0 P_0 + x_1 P_1 + r(\cos \theta \, P_2 + \sin \theta \, P_3)]\exp[\log z \, D] \ , &
\\ \vphantom{\frac{z}{\eta}}
\mathfrak{h} & = \{D, P_1, M_{01}, P_0 \} \ , &
f' & = \exp[\frac{r}{z}(\cos\theta \, P_2 + \sin \theta \, P_3)] \ , &
\\ \vphantom{\frac{z}{\eta}}
\bar{\mathfrak{h}} & = \{D, K_1, M_{01}, K_0\} \ , &
v & = \frac{x_1}\eta D + \frac{z}{2\eta} K_1 - \frac{x_0}{\eta} M_{01} \ . &
\end{flalign*}

\begin{flalign*}
5. \quad \vphantom{\frac{z}{\eta}}
r & = \eta \, D \wedge P_1 \ , &
f & = \exp[-t \cosh \chi \, P_0 + x_1 P_1 + t \sinh \chi \, (\cos \theta \, P_2 + \sin \theta \, P_3) ] \exp[\log z \, D] \ , &
\\ \vphantom{\frac{z}{\eta}}
\mathfrak{h} & = \{D, P_1 \} \ , &
f' & = \exp[\frac{t}{z}(-\cosh \chi \, P_0 + \sinh \chi \, (\cos \theta \, P_2 + \sin \theta \, P_3) ] \ ,
\\ \vphantom{\frac{z}{\eta}}
\bar{\mathfrak{h}} & = \{D, K_1\} \ , &
v & = \frac{x_1}\eta D + \frac{z}{2\eta} K_1 \ . &
\end{flalign*}

\begin{flalign*}
6. \quad \vphantom{\frac{z}{\eta}}
r & = \eta \, ( D \wedge P_2 + M_{23} \wedge P_3) \ , &
f & = \exp[t (-\cosh \chi \, P_0 + \sinh \chi \, P_1) + x_2 P_2 + x_3 P_3]\exp[\log z \, D] \ , &
\\ \vphantom{\frac{z}{\eta}}
\mathfrak{h} & = \{D, P_2, M_{23}, P_3\} \ , &
f' & = \exp[\frac{t}{z}(-\cosh \chi \, P_0 + \sinh \chi \, P_1)] \ , &
\\ \vphantom{\frac{z}{\eta}}
\bar{\mathfrak{h}} & = \{D, K_2, M_{23}, K_3\} \ , &
v & = \frac{x_2}\eta D + \frac{z}{2\eta} K_2 + \frac{x_3}{\eta} M_{23} \ . &
\end{flalign*}

\begin{flalign*}
7. \quad \vphantom{\frac{z}{\eta}}
r & = \eta \, ( M_{01} \wedge M_{+2} + M_{23} \wedge M_{+3}) \ , &
f & = \exp[-\tfrac12(x_- + \tfrac14 x_+(\psi_2^2 + \psi_3^2)) P_+ -\tfrac12 x_+ (P_- + \psi_2 P_2 + \psi_3 P_3)]\exp[\log z \, D] \ , &
\\ \vphantom{\frac{z}{\eta}}
\mathfrak{h} & = \{M_{01}, M_{+2}, M_{23}, M_{+3}\} \ , &
f' & = \exp[\sqrt{x_-x_+}P_0] \exp[\log z \, D] \ , &
\\ \vphantom{\frac{z}{\eta}}
\bar{\mathfrak{h}} & = \{M_{01}, M_{-2}, M_{23}, M_{-3}\} \ , &
v & = \frac{\psi_2}{2\eta} M_{01} + \frac{1}{2\eta} \sqrt{\frac{x_-}{x_+}} M_{-2} + \frac{\psi_3}{2\eta} M_{23} \ . &
\end{flalign*}

\begin{flalign*}
8. \quad \vphantom{\frac{z}{\eta}}
r & = \eta \, ( (D + M_{01}) \wedge P_+ + 2M_{+2} \wedge P_2 + 2M_{+3} \wedge P_3) \ , &
f & = \exp[-\ha(x_- P_+ + x_+ P_-) + r (\cos \theta \, P_2 + \sin \theta \, P_3)]\exp[\log z \, D] \ , &
\\ \vphantom{\frac{z}{\eta}}
\mathfrak{h} & = \{D+ M_{01} ,P_+ , M_{+2}, P_2 , M_{+3}, P_3\} \ , &
f' & = \exp[-\frac{x_+}{2} P_-] \ , &
\\ \vphantom{\frac{z}{\eta}}
\bar{\mathfrak{h}} & = \{D +M_{01}, K_-, M_{-2}, K_2, M_{-3}, K_3\} \ , &
v & = -\frac{x_-}{4\eta} (D + M_{01}) + \frac{z^2}{8\eta} K_- + \frac{z r}{4\eta}(\cos \theta \, M_{-2} + \sin \theta \, M_{-3}) \ . &
\end{flalign*}

\begin{flalign*}
9. \quad \vphantom{\frac{z}{\eta}}
r & = \eta \, ( (D + M_{01}) \wedge P_+ + 2M_{+2} \wedge P_2) \ , &
f & = \exp[-\ha(x_- P_+ + x_+ P_-) + r (\cos \theta \, P_2 + \sin \theta \, P_3)]\exp[\log z \, D] \ , &
\\ \vphantom{\frac{z}{\eta}}
\mathfrak{h} & = \{D+ M_{01} ,P_+ , M_{+2}, P_2 \} \ , &
f' & = \exp[-\frac{x_+}{2} P_- + \frac{r}{z} \sin\theta \, P_3] \ , &
\\ \vphantom{\frac{z}{\eta}}
\bar{\mathfrak{h}} & = \{D +M_{01}, K_-, M_{-2}, K_2\} \ , &
v & = -\frac{x_-}{4\eta} (D + M_{01}) + \frac{z^2}{8\eta} K_- + \frac{z r}{4\eta}\cos \theta \, M_{-2} \ . &
\end{flalign*}

\begin{flalign*}
10. \quad \vphantom{\frac{z}{\eta}}
r & = \eta \, (D + M_{01}) \wedge P_+ \ , &
f & = \exp[-\ha(x_- P_+ + x_+ P_-) + r (\cos \theta \, P_2 + \sin \theta \, P_3)]\exp[\log z \, D] \ , &
\\ \vphantom{\frac{z}{\eta}}
\mathfrak{h} & = \{D+ M_{01} ,P_+ \} \ , &
f' & = \exp[-\frac{x_+}{2} P_- + \frac{r}{z}(\cos \theta \, P_2 + \sin\theta \, P_3)] \ , &
\\ \vphantom{\frac{z}{\eta}}
\bar{\mathfrak{h}} & = \{D +M_{01}, K_-\} \ , &
v & = - \frac{x_-}{4\eta} (D + M_{01}) + \frac{z^2}{8\eta} K_- \ . &
\end{flalign*}

\begin{flalign*}
11. \quad \vphantom{\frac{z}{\eta}}
r & = \eta \, ( D \wedge P_+ + M_{+2} \wedge P_2 + M_{+3} \wedge P_3) \ , &
f & = \exp[-\ha(x_- P_+ + x_+ P_-) + r (\cos \theta \, P_2 + \sin \theta \, P_3)]\exp[\log z \, D] \ , &
\\ \vphantom{\frac{z}{\eta}}
\mathfrak{h} & = \{D ,P_+ , M_{+2}, P_2 , M_{+3}, P_3\} \ , &
f' & = \exp[-\frac{x_+}{2z} P_-] \ , &
\\ \vphantom{\frac{z}{\eta}}
\bar{\mathfrak{h}} & = \{D , K_-, M_{-2}, K_2, M_{-3}, K_3\} \ , &
v & = -\frac{x_-}{2\eta} D + \frac{z}{4\eta} K_- + \frac{r}{2\eta}(\cos \theta \, M_{-2} + \sin \theta \, M_{-3}) \ . &
\end{flalign*}

\begin{flalign*}
12. \quad \vphantom{\frac{z}{\eta}}
r & = \eta \, ( D \wedge P_+ + M_{+2} \wedge P_2) \ , &
f & = \exp[-\ha(x_- P_+ + x_+ P_-) + r (\cos \theta \, P_2 + \sin \theta \, P_3)]\exp[\log z \, D] \ , &
\\ \vphantom{\frac{z}{\eta}}
\mathfrak{h} & = \{D ,P_+ , M_{+2}, P_2 \} \ , &
f' & = \exp[-\frac{x_+}{2z} P_- + \frac{r}{z} \sin\theta \, P_3] \ , &
\\ \vphantom{\frac{z}{\eta}}
\bar{\mathfrak{h}} & = \{D, K_-, M_{-2}, K_2\} \ , &
v & = -\frac{x_-}{2\eta} D + \frac{z}{4\eta} K_- + \frac{r}{2\eta}\cos \theta \, M_{-2} \ . &
\end{flalign*}

\begin{flalign*}
13. \quad \vphantom{\frac{z}{\eta}}
r & = \eta \, D \wedge P_+ \ , &
f & = \exp[-\ha(x_- P_+ + x_+ P_-) + r (\cos \theta \, P_2 + \sin \theta \, P_3)]\exp[\log z \, D] \ , &
\\ \vphantom{\frac{z}{\eta}}
\mathfrak{h} & = \{D ,P_+ \} \ , &
f' & = \exp[-\frac{x_+}{2z} P_- + \frac{r}{z}(\cos \theta \, P_2 + \sin\theta \, P_3)] \ , &
\\ \vphantom{\frac{z}{\eta}}
\bar{\mathfrak{h}} & = \{D , K_-\} \ , &
v & = - \frac{x_-}{2\eta} D+ \frac{z}{4\eta} K_- \ . &
\end{flalign*}

\begin{flalign*}
14. \quad \vphantom{\frac{z}{\eta}}
r & = \eta \, ( M_{01} \wedge P_+ + M_{+2} \wedge P_2 + M_{+3} \wedge P_3) \ , &
f & = \exp[-\ha(x_- P_+ + x_+ P_-) + r (\cos \theta \, P_2 + \sin \theta \, P_3)]\exp[\log z \, D] \ , &
\\ \vphantom{\frac{z}{\eta}}
\mathfrak{h} & = \{M_{01} ,P_+ , M_{+2}, P_2 , M_{+3}, P_3\} \ , &
f' & = \exp[-\frac{x_+}{2z} P_-] \exp[\log z \, D] \ , &
\\ \vphantom{\frac{z}{\eta}}
\bar{\mathfrak{h}} & = \{M_{01} , K_-, M_{-2}, K_2, M_{-3}, K_3\} \ , &
v & = -\frac{x_-}{2\eta} M_{01}+ \frac{1}{4\eta z} K_- + \frac{r}{2\eta z}(\cos \theta \, M_{-2} + \sin \theta \, M_{-3}) \ . &
\end{flalign*}

\begin{flalign*}
15. \quad \vphantom{\frac{z}{\eta}}
r & = \eta \, ( M_{01} \wedge P_+ + M_{+2} \wedge P_2) \ , &
f & = \exp[-\ha(x_- P_+ + x_+ P_-) + r (\cos \theta \, P_2 + \sin \theta \, P_3)]\exp[\log z \, D] \ , &
\\ \vphantom{\frac{z}{\eta}}
\mathfrak{h} & = \{M_{01} ,P_+ , M_{+2}, P_2 \} \ , &
f' & = \exp[-\frac{x_+}{2z} P_- + r \sin\theta \, P_3]\exp[\log z \, D] \ , &
\\ \vphantom{\frac{z}{\eta}}
\bar{\mathfrak{h}} & = \{M_{01}, K_-, M_{-2}, K_2\} \ , &
v & = -\frac{x_-}{2\eta} M_{01} + \frac{1}{4\eta z} K_- + \frac{r}{2\eta z}\cos \theta \, M_{-2} \ . &
\end{flalign*}

\begin{flalign*}
16. \quad \vphantom{\frac{z}{\eta}}
r & = \eta \, M_{01} \wedge P_+ \ , &
f & = \exp[-\ha(x_- P_+ + x_+ P_-) + r (\cos \theta \, P_2 + \sin \theta \, P_3)]\exp[\log z \, D] \ , &
\\ \vphantom{\frac{z}{\eta}}
\mathfrak{h} & = \{M_{01} ,P_+ \} \ , &
f' & = \exp[-\frac{x_+}{2z} P_- + r(\cos \theta \, P_2 + \sin\theta \, P_3)]\exp[\log z \, D] \ , &
\\ \vphantom{\frac{z}{\eta}}
\bar{\mathfrak{h}} & = \{M_{01} , K_-\} \ , &
v & = -\frac{x_-}{2\eta} M_{01}+ \frac{1}{4\eta z} K_- \ . &
\end{flalign*}

\begin{flalign*}
17. \quad \vphantom{\frac{z}{\eta}}
r & = \eta \, M_{01} \wedge (P_+ + K_+) \ , &
f & = \exp[-\ha(x_- P_+ + x_+ P_-) + r (\cos \theta \, P_2 + \sin \theta \, P_3)]\exp[\log z \, D] \ , &
\\ \vphantom{\frac{z}{\eta}}
\mathfrak{h} & = \{M_{01} ,P_+ + K_+ \} \ , &
f' & = \exp[\sqrt{x_-x_+} P_0 + r(\cos \theta \, P_2 + \sin\theta \, P_3)]\exp[\log z \, D] \ , &
\\ \vphantom{\frac{z}{\eta}}
\bar{\mathfrak{h}} & = \{M_{01} , P_- + K_-\} \ , &
v & = \frac{1}{8\eta}\sqrt{\frac{x_-}{x_+}} (P_- + K_-) \ . &
\end{flalign*}

\begin{flalign*}
18. \quad \vphantom{\frac{z}{\eta}}
r & = \eta \, M_{01} \wedge (P_+ - K_+) \ , &
f & = \exp[-\ha(x_- P_+ + x_+ P_-) + r (\cos \theta \, P_2 + \sin \theta \, P_3)]\exp[\log z \, D] \ , &
\\ \vphantom{\frac{z}{\eta}}
\mathfrak{h} & = \{M_{01} ,P_+ - K_+ \} \ , &
f' & = \exp[\sqrt{x_-x_+} P_0 + r(\cos \theta \, P_2 + \sin\theta \, P_3)]\exp[\log z \, D] \ , &
\\ \vphantom{\frac{z}{\eta}}
\bar{\mathfrak{h}} & = \{M_{01} , P_- - K_-\} \ , &
v & = -\frac{1}{8\eta}\sqrt{\frac{x_-}{x_+}} (P_- - K_-) \ . &
\end{flalign*}

\begin{flalign*}
19. \quad \vphantom{\frac{z}{\eta}}
r & = \eta \, (D-\zeta P_1) \wedge P_0 \ , &
f & = \exp[-x_0 P_0 + x_1 P_1 + r (\cos \theta \, P_2 + \sin \theta \, P_3)]\exp[\log z \, D] \ , &
\\ \vphantom{\frac{z}{\eta}}
\mathfrak{h} & = \{D-\zeta P_1,P_0\} \ , &
f' & = \exp[(\frac{x_1 + \zeta z - \zeta}z) P_1 + \frac{r}{z}(\cos \theta \, P_2 + \sin \theta \, P_3)] \ , &
\\ \vphantom{\frac{z}{\eta}}
\bar{\mathfrak{h}} & = \{D - \zeta K_1, K_0\} \ , &
v & = -\frac{x_0}{\eta(1+2\zeta^2)} (D-\zeta K_1) + \frac{z}{2\eta} K_0 \ . &
\end{flalign*}

\begin{flalign*}
20. \quad \vphantom{\frac{z}{\eta}}
r & = \eta \, (D + M_{01} - \zeta M_{23}) \wedge P_+ \ , &
f & = \exp[-\ha(x_- P_+ + x_+ P_-) + r (\cos \theta \, P_2 + \sin \theta \, P_3)]\exp[\log z \, D] \ , &
\\ \vphantom{\frac{z}{\eta}}
\mathfrak{h} & = \{D+ M_{01} - \zeta M_{23} ,P_+ \} \ , &
f' & = \exp[-\frac{x_+}{2} P_- + \frac{r}{z}(\cos (\theta \, - \zeta \log z) P_2 + \sin(\theta \, - \zeta \log z) P_3)] \ , &
\\ \vphantom{\frac{z}{\eta}}
\bar{\mathfrak{h}} & = \{D +M_{01} + \zeta M_{23}, K_-\} \ , &
v & = - \frac{x_-}{2\eta(2+\zeta^2)} (D + M_{01} + \zeta M_{23}) + \frac{z^2}{8\eta} K_- \ . &
\end{flalign*}
\end{changemargin}

\normalsize

%%%%%%%%%%%%%%%%%%%%%%%%%%%%%%%%%%%%%%

%%%%%%%%%%%%%%%%%%%%%%%%%%%%%%%%%%%%%%

%%%%%%%%%%%%%%%%%%%%%%%%%%%%%%%%%%%%%%
\end{document}
%%%%%%%%%%%%%%%%%%%%%%%%%%%%%%%%%%%%%%